\documentclass[aps,prd,twocolumn,showpacs,groupedaddress,amsmath,amssymb,nofootinbib]{revtex4}

\usepackage{graphicx}% Include figure files
\usepackage{dcolumn}% Align table columns on decimal point
\usepackage{bm}% bold math

\begin{document}
\title{Unitary gauge considered harmful}

\author{Tommy Anderberg}
\email{Tommy.Anderberg@simplicial.net}
%\thanks{}
%\affiliation{}
\noaffiliation

\date{\today}

\begin{abstract}
I informally review the Higgs mechanism, focusing on fundamental
aspects which should be common knowledge but apparently are not,
explain why your understanding of gauge symmetry breaking is a
decreasing function of your reliance on the unitary gauge, and
discuss some implications for cosmology (domains, dark energy, CMB
cold spot).
\end{abstract}

%11.15.Ex Spontaneous breaking of gauge symmetries
%11.27.+d Extended classical solutions; cosmic strings, domain walls, texture (see also 98.80.Cq in cosmology; 11.25.-w Strings and branes)
%11.30.Qc Spontaneous and radiative symmetry breaking
%11.30.Rd Chiral symmetries
%12.10.-g Unified field theories and models (see also 04.50.-h Higher-dimensional gravity and other theories of gravity-in general relativity and gravitation, 11.25.Mj Compactification and four-dimensional models)
%12.10.Dm Unified theories and models of strong and electroweak interactions
%12.10.Kt Unification of couplings; mass relations
%12.15.Ji Applications of electroweak models to specific processes
%95.30.Cq Elementary particle processes
%95.36.+x Dark energy (see also 98.80.-k Cosmology)
%97.10.Xq Luminosity and mass functions
%98.80.-k Cosmology (see also section 04 General relativity and gravitation; for origin and evolution of galaxies, see 98.62.Ai; for elementary particle and nuclear processes, see 95.30.Cq; for dark matter, see 95.35.+d; for dark energy, see 95.36.+x; for superclusters and large-scale structure of the Universe, see 98.65.Dx)
%98.80.Cq Particle-theory and field-theory models of the early Universe (including cosmic pancakes, cosmic strings, chaotic phenomena, inflationary universe, etc.)
%\keywords{GUT, electroweak, spontaneous symmetry breaking, texture, CMB, cold spot}

\maketitle

%\subsection{}
%\subsubsection{}

\section{Introduction}
Long ago, Dijkstra observed that ``the quality of programmers is a
decreasing function of the density of go to statements in the
programs they produce'' and famously argued that gotos should be
abolished from high level programming languages \cite{dijkstra1968}.
If you are old (or geeky) enough to have ever used gotos, you know
why: while they can simplify the task at hand and even make programs
run faster, their repeated use quickly leads to ``spaghetti code''.
That's a technical term for ``incomprehensible mess''. The short
term efficiency gain is paid for with a loss of understanding, which
inevitably ends up causing errors.

In recent years, I have observed an analogous phenomenon in physics.
Initially, I thought that it only occurred in isolated and
particularly unfortunate cases (after all, finding oneself talking
to me could be considered pretty unfortunate). As my sample has
grown to include distinguished practitioners (or so I'm told), I
have reluctantly been forced to accept that it is not so. The
problem is ubiquitous.

With only months remaining before \$10 billion worth of LHC start
looking for the Higgs (or whatever actually breaks electroweak
symmetry) it is my regrettable duty to inform you that many,
probably even most of you do not understand gauge symmetry breaking.
The culprit is none other than the ``goto'' of particle physics: the
unitary gauge. Like the goto statement, the unitary gauge does
simplify some tasks (proving unitarity, estimating mass spectra,
introducing students to gauge symmetry breaking), but only at the
cost of understanding. This loss of understanding is causing errors,
some of them potentially spectacular.

\section{Classical theory}

\subsection{Goldstone model}
The standard textbook example of spontaneous symmetry breaking (SSB)
is the Goldstone model \cite{goldstone1961}, defined by the Lagrangian
\begin{eqnarray}\label{goldstone}
{\cal L}(\phi) = (\partial_\mu\phi)^\dag(\partial^\mu\phi) - V(\phi)
\end{eqnarray}
with ``Mexican hat'' potential
\begin{eqnarray}\label{mexPotential}
V(\phi) = \lambda\left[\phi^\dag\phi - \frac{\nu^2}{2}\right]^2
\end{eqnarray}
where $\phi = \phi(x)$ is a complex Lorentz scalar field and
$\lambda$, $\nu$ are real, positive parameters. $V(\phi)$ has a
degenerate ``valley'' of connected minima at $\phi^\dag\phi = \nu^2/2$,
i.e. a circle in the complex plane, and a local maximum at
$\phi^\dag\phi = 0$.

Solving the full equations of motion may be too hard for us, but we
can at least obtain approximate solutions using perturbation theory.
To do so, we must pick a stable point $\langle\phi\rangle$ about
which small oscillations in $\phi$ will remain small. That means a
point in the valley. Since gradient energy is minimized when
$\phi(x)$ has the same value for all spacetime coordinates $x$, any
constant $\phi$ in the valley minimizes total energy and therefore
qualifies as a ground state (``vacuum'') of ${\cal L}(\phi)$. The
circle $\phi^\dag\phi = \nu^2/2$ is therefore known as the vacuum
manifold of ${\cal L}(\phi)$. It has a global U(1) $\sim$ O(2)
symmetry, i.e. it's invariant under rotations of $\phi$ in the
complex plane. Picking a particular vacuum breaks this symmetry.
This is the essence of SSB: the theory (meaning the Lagrangian), has
more symmetry than any individual ground state.

Let's arbitrarily choose the point
\begin{eqnarray}\label{aGoldstoneVacuum}
\langle\phi\rangle = \frac{\nu}{\sqrt{2}}
\end{eqnarray}
on the real axis as the starting point for our perturbative
expansion. To parameterize perturbations about it, introduce
orthogonal field coordinates $\phi_1$
(along the real axis) and $\phi_2$ (along the imaginary axis):
\begin{eqnarray}\label{cartesianPhi}
\phi &=& \langle\phi\rangle + \frac{\phi_1 + i \phi_2}{\sqrt{2}} \nonumber\\
&=& \frac{\nu + \phi_1 + i \phi_2}{\sqrt{2}}
\end{eqnarray}
In terms of $\phi_1$ and $\phi_2$, the Lagrangian is
\begin{eqnarray}\label{cartesianGoldstone}
{\cal L}(\phi_1, \phi_2) = \frac{1}{2}(\partial_\mu\phi_1)^2 +
\frac{1}{2}(\partial_\mu\phi_2)^2 - V(\phi_1, \phi_2)
\end{eqnarray}
with potential
\begin{eqnarray}\label{cartesianMexPotential}
V(\phi_1, \phi_2) &=& \frac{\lambda}{4}\left[\phi_1^2 + \phi_2^2 +
2\nu\phi_1\right]^2 \nonumber\\
&=& \lambda\nu^2\phi_1^2 + \lambda\nu\phi_1(\phi_1^2 + \phi_2^2) \nonumber\\
&&+\thinspace\lambda\left(\frac{\phi_1^2 + \phi_2^2}{2}\right)^2
\end{eqnarray}
Note the first term: it gives $\phi_1$ a mass squared
$2\lambda\nu^2$. There is no term quadratic in $\phi_2$, so $\phi_2$
is massless.

The reason for this difference is that $\phi_1$ describes
displacements against the restoring force of the potential, while
$\phi_2$ describes displacements along the flat potential valley.
The generalization of this observation is Goldstone's theorem:
spontaneous breaking of a continuous symmetry implies the existence
of a massless field (a Goldstone boson; in supersymmetric theories,
there are also Goldstone fermions -- goldstinos -- to match) for
each flat direction. In mathematical terms, the number of such
Goldstone bosons is the dimension of the coset space G/H, where G
is the full symmetry group of the vacuum manifold and H is the
subgroup of G under which the vacuum remains invariant (if any).

The parametrization of Eq. \eqref{cartesianPhi} using Cartesian
field coordinates $\phi_1$ and $\phi_2$ is the natural choice for a
perturbative expansion, but ill suited to exploring ${\cal L}$
beyond the infinitesimal neighborhood of a particular vacuum, i.e.
to finding non-perturbative solutions. For that, it's more
convenient to use curvilinear coordinates which follow the geometry
of the vacuum manifold. In the simple case of the Goldstone model,
this means standard polar coordinates:
\begin{eqnarray}\label{nlPhi}
\phi &=& \left(\langle\phi\rangle + \frac{\rho}{\sqrt{2}}\right) \textrm{e}^{i \theta}\nonumber\\
&=& \frac{\nu + \rho}{\sqrt{2}} \textrm{e}^{i \theta}
\end{eqnarray}
The leading terms of a Taylor expansion of $\phi$ in the radial and
angular displacements $\rho$ and $\theta$ are
\begin{eqnarray}\label{nlPhiTaylor}
\phi &=& \frac{\nu + \rho + i \nu\theta}{\sqrt{2}} + ...
\end{eqnarray}
By comparison with Eq. \eqref{cartesianPhi}, for small perturbations
\begin{eqnarray}
\phi_1 &\simeq& \rho \label{nlPhi1Equivalence} \\
\phi_2 &\simeq& \nu\theta \label{nlPhi2Equivalence}
\end{eqnarray}
($\phi_2$ parameterizes the tangent of the vacuum manifold at $\phi
= \langle\phi\rangle$). In terms of $\rho$ and $\theta$, the Lagrangian is
\begin{eqnarray}\label{nlGoldstone}
{\cal L}(\rho, \theta) &=&
\frac{1}{2}(\partial_\mu\rho)(\partial^\mu\rho) + \frac{1}{2}(\rho +
\nu)^2 (\partial_\mu\theta)(\partial^\mu\theta) \nonumber \\
&&-\thinspace V(\rho)
\end{eqnarray}
with $\theta$-independent potential
\begin{eqnarray}\label{nlMexPotential}
V(\rho) &=& \lambda \left[\rho (\nu + \rho/2)\right]^2 \nonumber\\
&=& \lambda\left[\frac{\rho^4}{4} + \nu \rho^3 + \nu^2 \rho^2\right]
\end{eqnarray}
from which we again read off a mass squared $2\lambda\nu^2$ for the
radial excitation ($\rho$), while $\theta$ remains massless.

\subsection{Higgs mechanism}
The Goldstone model is turned into the simplest example of the Higgs
mechanism \cite{higgs1964}\cite{higgs1966} by upgrading its global
U(1) symmetry to a local U(1) symmetry. The Goldstone Lagrangian is
invariant under an identical rotation of $\phi(x)$ at every
spacetime point $x$; the Higgs Lagrangian is invariant under an
independent rotation at each spacetime point.

To construct it, introduce an Abelian gauge field $A_\mu(x)$ with
field strength tensor
\begin{eqnarray}\label{u1FieldStrength}
F_{\mu\nu} = \partial_\mu A_\nu - \partial_\nu A_\mu
\end{eqnarray}
and Lagrangian
\begin{eqnarray}\label{u1GaugeLagrangian}
{\cal L}_A = -\frac{1}{4} F_{\mu\nu}F^{\mu\nu}
\end{eqnarray}
and substitute ordinary derivatives with covariant derivatives
\begin{eqnarray}\label{u1CovariantDerivative}
\partial_\mu \to D_\mu = \partial_\mu + i g A_\mu
\end{eqnarray}
($g$ is the gauge coupling constant) in ${\cal L}(\phi)$. The
resulting total Lagrangian is invariant under the simultaneous local
transformations
\begin{eqnarray}
\phi(x) \to \phi'(x) &=& \textrm{e}^{-i\omega(x)} \phi(x)
\label{u1PhiGaugeTransformation} \\
A_\mu(x) \to A'_\mu(x) &=& A_\mu(x) + \frac{1}{g}\partial_\mu
\omega(x) \label{u1AGaugeTransformation}
\end{eqnarray}
The variation in the gauge field $A_\mu(x)$ exactly compensates the
variation in $\phi(x)$, i.e. if we apply the transformations of Eqs.
\eqref{u1PhiGaugeTransformation} and \eqref{u1AGaugeTransformation}
to the total Lagrangian and work through the algebra, we end up with
exactly the same expression, apart from the trivial substitutions
\begin{eqnarray}
\phi(x) &\to& \phi'(x) \label{u1PhiSubstitution} \\
A_\mu(x) &\to& A'_\mu(x) \label{u1ASubstitution}
\end{eqnarray}
This freedom to apply a local transformation implies that there are
fewer independent equations of motion than degrees of freedom. The
problem is of course the gauge field. Its Lagrangian is identical to
that of a photon, with $\phi$ in the role of a charged scalar, so
what we have here is just scalar electrodynamics with an unusual
$V(\phi)$. We know that photons have only two independent degrees of
freedom, the transverse polarization states, but $A_\mu(x)$ has
four. As in electrodynamics, we must therefore supplement the
equations of motion with a gauge fixing condition before we can
actually compute anything.

Following the standard textbook approach, we postpone that decision and
choose a vacuum first. Choosing again
$\langle\phi\rangle = \nu/\sqrt{2}$ and expanding
$(\partial_\mu\phi)^\dag(\partial^\mu\phi) \to
(D_\mu\phi)^\dag(D^\mu\phi)$ yields
\begin{eqnarray}\label{u1DerivativeTerms}
(D_\mu\phi)^\dag(D^\mu\phi) &=& (\partial_\mu
\phi)^\dag(\partial^\mu
\phi) \nonumber \\
&&+ i g A_\mu ((\partial^\mu \phi)^\dag \phi - \phi^\dag
(\partial^\mu\phi)) \nonumber \\
&&+ g^2 A_\mu A^\mu \phi^\dag \phi
\end{eqnarray}
In Cartesian coordinates, the new terms are
\begin{eqnarray}\label{newCartesianHiggsTerms}
(g\nu)^2 A_\mu A^\mu \nonumber \\
+ g A_\mu (\phi_1 \partial^\mu\phi_2 - \phi_2 \partial^\mu\phi_1) \nonumber \\
+ g^2 A_\mu A^\mu (\nu\phi_1 + (\phi_1^2 + \phi_2^2)/2) \nonumber \\
+ g \nu A_\mu (\partial^\mu \phi_2)
\end{eqnarray}
In polar coordinates, they are
\begin{eqnarray}\label{newPolarHiggsTerms}
(g\nu)^2 A_\mu A^\mu \nonumber \\
+ g A_\mu \rho (\partial^\mu \theta)(2\nu + \rho) \nonumber \\
+ g^2 A_\mu A^\mu \rho (2\nu + \rho)/2 \nonumber \\
+ g \nu^2 A_\mu \partial^\mu \theta
\end{eqnarray}
The first term tells us that the gauge field $A_\mu$ has picked up a
mass squared $2(g\nu)^2$. The remaining terms, except for the last
one, describe interactions involving $A_\mu$, $\phi_1 \sim \rho$
(massive as in the Goldstone case) and $\phi_2 \sim \nu\theta$
(still massless).

The last term is problematic: it's quadratic in the fields, like a
mass term, but it mixes $A_\mu$ and $\phi_2 \sim \nu\theta$,
suggesting that they are not independent. Counting degrees of
freedom confirms this. We started out with a complex scalar (two
real components) and a massless vector field (two more) for a total
of four degrees of freedom. We ended up with two scalars and a
massive vector field, which also has a longitudinal polarization
state, for a total of 2 + 3 = 5 degrees of freedom. Since a simple
change of coordinates can not affect the number of degrees of
freedom, one of them is redundant.

This is where your favorite introductory field theory textbook
points out that $\phi_2 \sim \nu\theta$ can be made to vanish at
every spacetime point $x$ using the invariance of the total
Lagrangian under the simultaneous local transformations of Eqs.
\eqref{u1PhiGaugeTransformation} and \eqref{u1AGaugeTransformation}.
In polar coordinates, this is trivially easy to see: simply set the
transformation parameter $\omega(x)$ equal to the angular
displacement $\theta(x)$. The transformed field $\phi'(x)$ will then
be real, i.e. $\phi'_2 \sim \nu\theta' = 0$:
\begin{eqnarray}\label{u1UnitaryPhi}
\phi'(x) &=& \textrm{e}^{-i\theta(x)} \phi(x)\nonumber\\
&=& \textrm{e}^{-i\theta(x)}\frac{\nu + \rho(x)}{\sqrt{2}}
\textrm{e}^{{i \theta(x)}}\nonumber\\
&=&\frac{\nu + \rho(x)}{\sqrt{2}}
\end{eqnarray}
Provided that we also perform the transformation
\begin{eqnarray}\label{u1AUnitaryA}
A_\mu(x) \to A'_\mu(x) = A_\mu(x) + \frac{1}{g}\partial_\mu
\theta(x)
\end{eqnarray}
the only effect on the Lagrangian, expressed in terms of $\phi$ and
$A_\mu$, is to adorn the fields with primes according to Eqs.
\eqref{u1PhiSubstitution}-\eqref{u1ASubstitution}.

We conclude that gauge invariance allows us to impose the condition
\begin{eqnarray}\label{u1UnitaryGauge}
\theta(x) = 0 \thinspace \Leftrightarrow \thinspace \phi_2(x) = 0
\end{eqnarray}
so as to remove the extra degree of freedom (known as a ``would-be''
Goldstone boson) and leave us with a Lagrangian written exclusively
in terms of a massive $A_\mu(x)$ (colloquially said to have
``eaten'' the would-be Goldstone boson in order to acquire a
longitudinal component) and a massive $\phi_1(x) \simeq \rho(x)$.
The latter is the simplest example of a Higgs boson. The condition
of Eq. \eqref{u1UnitaryGauge} (and its equivalents for larger
symmetry groups) is the unitary gauge.

\subsection{Can (maybe), not must!}
The fact that we \emph{can} impose the unitary gauge does not imply
that we \emph{must} do so, of course. There is a literally infinite
number of conditions which may legitimately be used to remove the
redundant degree of freedom, and you are free to use whichever is
most convenient. Physical observables like energy density are by
definition independent of this choice, but if you wish to directly
compare solutions obtained in different gauges, all you have to do
is transform them to a common gauge.

The main requirement on a gauge condition is that it be reversible:
given an arbitrary field configuration $(A_\mu(x), \phi(x))$, a
transformation must exist such that the transformed fields satisfy the
condition \emph{and} can be uniquely transformed back to the original
configuration (see e.g. p. 7 in \cite{faddeev1991}). This ensures that no
information is lost by applying the condition. If this requirement is not
satisfied, strictly speaking the condition is not a gauge condition, but
an unphysical constraint. It may still be useful in special situations,
if it only cuts out a part of configuration space disjoint from that of
the configurations under consideration, but it can not be used in full
generality.

It is not hard to see that the unitary ``gauge'' is in fact such an
unphysical constraint: it fails to be reversible at $\phi(x) = 0$.

A famous example of the solutions living in the null space of the
unitary gauge (its ``blind spot'') is provided by the topologically
stable vortices first described by Nielsen and Olesen
\cite{nielsen1973}. Such a vortex is characterized by a non-trivial
map from polar field coordinates ($\rho$, $\theta$) to polar spatial
coordinates ($r$, $\varphi$)
\begin{eqnarray}\label{u1VortexPhi}
\phi(r, \varphi) = \rho(r)\textrm{e}^{i n \varphi}
\end{eqnarray}
and satisfies
\begin{eqnarray}\label{u1VortexRhoLimits}
\lim\limits_{r \to \infty} \partial_r \rho(r) &=& 0 \\
\lim\limits_{r \to 0} \rho(r) &=& 0
\end{eqnarray}
Extend it to a cylinder and you get a flux tube (or ``string'', not
to be confused with fundamental ones) known to cosmologists as a
cosmic string. The ``winding number'' $n$ is an integer (the
Pontryagin index) specifying the number of times $\phi$ goes around
the potential valley as you walk along a single loop about spatial
origin. The stability argument is simple: changing $n$ requires the
field to be lifted out of the potential valley and slid over the top
of the Mexican hat, at an energy density cost $\propto
\lambda\nu^4$.

If you try to transform a Nielsen-Olesen vortex to the unitary gauge
throughout all space, you will inevitably run into trouble with Eq.
\eqref{u1AUnitaryA} at $r = 0$: $\theta$ is undefined there, so
$A'_\mu$ will be undefined too. The singularity at $\phi(x) = 0$
blinds the unitary gauge to this kind of solution. Nielsen and
Olesen instead used the time-axial gauge $A_0 = 0$, and in so doing
established the gauge of choice for non-perturbative work in gauge
field theories with SSB
\cite{dashen1974}\cite{klinkhamer1984}\cite{ambjorn1991}\cite{galtsov1991}\cite{biro1993}\cite{rajantie2001}\cite{tranberg2003}\cite{graham2007}\cite{diaz-gil2008}
\footnote{It would be interesting to compare notes with these
authors. Were they too, at some point, summarily dismissed with
claims that ``the true dynamics is given by the unitary gauge''?}.

The lesson here is that while SSB is an intrinsically
non-perturbative phenomenon, the unitary gauge is \emph{all} about
perturbation theory. There is nothing wrong with the Higgs
Lagrangian expressed in terms of the original $\phi$ field, i.e.
before shifting field coordinates to a particular vacuum. The
shift's only purpose is to allow the use of perturbation theory. If
you are not after a perturbative expansion, skip the shift and use a
standard gauge condition for the gauge field(s). If losing the
convenience of manifest Lorentz covariance is not a concern, a
simple condition like the radiation gauge $\nabla\cdot\textbf{A} =
0$ or an axial gauge is all you need.

\subsection{Non-Abelian Higgs mechanism}
There are two ways to summarize the construction of the Abelian
Higgs model. You could say that we take the Goldstone model and
gauge its U(1) symmetry (as we did above) or you could say that
we take an Abelian gauge field and ``Higgs'' it. Either way,
the procedure is readily generalized to the non-Abelian case.

Consider a generic Yang-Mills (i.e. non-Abelian) gauge field theory
\cite{yang1954} (henceforth simply YM), defined by the Lagrangian
\begin{eqnarray}\label{nonAbelianGaugeLagrangian}
{\cal L}_W = -\frac{1}{4} F_{a\mu\nu}F_a^{\mu\nu}
\end{eqnarray}
where the field strength tensor $F_{a\mu\nu}$ and gauge field
$W_{a\mu}$ now carry the group index $a$
\begin{eqnarray}\label{nonAbelianFieldStrength}
F_{a\mu\nu} = \partial_\mu W_{a\nu} - \partial_\nu W_{a\mu} +
g\thinspace C_{abc}W_{b\mu}W_{c\nu}
\end{eqnarray}
and $C_{abc}$ are the totally antisymmetric structure constants of
some simple Lie group, such that representation matrices ${\bm T}_a$
satisfy
\begin{eqnarray}\label{nonAbelianGroupStructure}
\left[{\bm T}_a, {\bm T}_b\right] = i\thinspace C_{abc} {\bm T}_c
\end{eqnarray}
To preserve gauge invariance, any matter field $\Psi(x)$ added to
the Lagrangian (fermion or scalar, at this point we don't care) and
coupling to $W_{a\mu}$ must transform according to
\begin{eqnarray}\label{nonAbelianPhiTransformation}
\Psi \to \Psi' = {\bm U}\thinspace\Psi
\end{eqnarray}
with
\begin{eqnarray}\label{nonAbelianU}
{\bm U}(x) = \textrm{e}^{-i\thinspace\omega_a(x)\thinspace{\bm T}_a}
\end{eqnarray}
and transformation parameters $\omega_a(x)$. The covariant
derivative
\begin{eqnarray}\label{nonAbelianCovariantDerivative}
{\bm D}_\mu = \partial_\mu + i g W_{a\mu} {\bm T}_a = \partial_\mu +
i g {\bm W}_\mu
\end{eqnarray}
compensates the variation in terms containing ${\bm D}_\mu \Psi$ (by
making them transform like $\Psi$, so that globally invariant
combinations like $\Psi^\dag \partial_\mu \Psi$ become locally
invariant after the substitution $\partial_\mu \to {\bm D}_\mu$)
provided that the gauge fields undergo the simultaneous
transformation
\begin{eqnarray}\label{nonAbelianWGaugeTransformation}
W_{a\mu}{\bm T}_a &\to& W'_{a\mu} {\bm T}_a \nonumber\\
&=& {\bm U}W_{a\mu}{\bm T}_a{\bm U}^\dag +
\frac{i}{g}\left(\partial_\mu {\bm U}\right){\bm U}^\dag
\end{eqnarray}
Note that we recover the Abelian case when ${\bm T} = 1$.

The smallest non-Abelian Lie group is SU(2), with structure
constants
\begin{eqnarray}\label{su2Cabc}
C_{abc} = \varepsilon_{abc}
\end{eqnarray}
(the totally antisymmetric Levi-Civita symbol, with
$\varepsilon_{123} = 1$). In the fundamental (spinorial)
representation,
\begin{eqnarray}\label{su2FundamentalTs}
{\bm T}_a = \frac{\tau_a}{2}
\end{eqnarray}
where $\tau_1$, $\tau_2$, $\tau_3$ are the Pauli matrices
\begin{eqnarray}\label{pauliMatrices}
\tau_1 = \left[
\begin{array}{lr}
0 & 1\\
1 & 0
\end{array}
\right]\quad \tau_2 = \left[
\begin{array}{lr}
0 & -i\\
i & 0
\end{array}
\right]\quad \tau_3 = \left[
\begin{array}{lr}
1 & 0\\
0 & -1
\end{array}
\right]
\end{eqnarray}
A scalar field in this representation must transform as an SU(2)
doublet, i.e.
\begin{eqnarray}\label{su2phi}
\Phi = \begin{bmatrix}
\phi^+\\
\phi^0
\end{bmatrix} \to \Phi' = \textrm{e}^{-\frac{i}{2}\omega_a\tau_a}\thinspace
\begin{bmatrix}
\phi^+\\
\phi^0
\end{bmatrix}
\end{eqnarray}
The superscripts ``+'' and ``0'' are just labels (for now).

To induce SSB, we simply substitute this $\Phi$ doublet into the
Goldstone Lagrangian, Eq. \eqref{goldstone}, let $\partial_\mu \to
{\bm D}_\mu$ and add the result to ${\cal L}_W$. In terms of the
real $\Phi$ components $\phi_1$, $\phi_2$, $\phi_3$ and $\phi_4$
\begin{eqnarray}\label{su2phiComponents}
\phi^+ &=& \phi_3 + i \phi_4 \\
\phi^0 &=& \phi_1 + i \phi_2
\end{eqnarray}
the (Higgs) vacuum manifold is now a 3-sphere defined by
\begin{eqnarray}\label{su2phiVacuumManifold}
\Phi^\dag\Phi = (\phi_1)^2 + (\phi_2)^2 + (\phi_3)^2 + (\phi_4)^2 =
\nu^2/2
\end{eqnarray}
There is actually more than we bargained for here: the symmetry of a
3-sphere is O(4) $\sim$ SU(2)$\times$SU(2), twice the size of the
SU(2) which we wish to break. This choice is made in anticipation of
quantization. Classically, any symmetric potential will do, but if
we want the quantized model to be renormalizable, the potential can
be at most quartic in the fields (see \cite{li1974} for a systematic
survey of SU(N) and O(N) symmetry breaking patterns).

To estimate the perturbative mass spectrum, repeat the steps
followed in the Abelian case. Shift the origin of our $\Phi$
coordinates to
\begin{eqnarray}\label{anSu2HiggsVacuum}
\langle\Phi\rangle = \frac{1}{\sqrt{2}}\begin{bmatrix}
0\\
\nu
\end{bmatrix}
\end{eqnarray}
(i.e. let $\phi_0 \to \nu/\sqrt{2} + \phi_0$), expand the $({\bm
D}_\mu \Phi)^\dagger ({\bm D}^\mu \Phi)$ term in the Higgs
Lagrangian and read off the quadratic terms:
\begin{eqnarray}\label{su2Wmass}
\frac{1}{2}\left(\frac{g\nu}{2}\right)^2 W_{a\mu} W_a^\mu +
\frac{\lambda\nu^2}{2} (\phi_0)^2
\end{eqnarray}
All three gauge bosons have become massive, along with the radial
$\Phi$ component in the chosen vacuum; but as expected, there are
also terms mixing gauge bosons and derivatives of the angular $\Phi$
components. They can again be removed by transforming to the unitary
gauge: pass to ``polar'' field coordinates
\begin{eqnarray}\label{su2PolarPhi}
\Phi(x) = \frac{1}{\sqrt{2}}
\textrm{e}^{\frac{i}{2}\thinspace\theta_a(x)\thinspace\tau_a}
\begin{bmatrix}
0\\
\nu + \rho(x)
\end{bmatrix}
\end{eqnarray}
and set the transformation parameters $\omega_a(x)$ of Eq.
\eqref{su2phi} equal to the angular displacement $\theta(x)$. Just
as in the Abelian case, the exponentials cancel, and we are left
with the real, radial component
\begin{eqnarray}\label{su2TransformedPhi}
\Phi'(x) = \frac{1}{\sqrt{2}}
\begin{bmatrix}
0\\
\nu + \rho(x)
\end{bmatrix}
\end{eqnarray}
The would-be Goldstone bosons are gone, ``eaten'' by the gauge bosons,
which have in turn been transformed according to Eq.
\eqref{nonAbelianWGaugeTransformation}.

It should now be evident by inspection of Eq. \eqref{nonAbelianU}
that the analogous procedure will work for any semisimple Lie group,
i.e. for any product of simple Lie groups. Historically, the
electroweak sector of the Standard Model
\cite{weinberg1967}\cite{salam1968} was constructed this way, by
Higgsing an existing gauge invariant Lagrangian, Glashow's
U(1)$\times$SU(2) model \cite{glashow1961}.

\subsection{Instantons and sphalerons}\label{section:instantonsandsphalerons}
As we have already seen, the unitary gauge does not tell the whole
story about the Abelian Higgs model. In the non-Abelian case, it
reveals even less.

Consider again the Nielsen-Olesen vortex of Eq. \eqref{u1VortexPhi}.
In topological terms, it is stable because a non-trivial map from
polar spatial coordinates to polar field coordinates, i.e. from a
circle (spatial infinity) to another circle (the vacuum manifold),
can not be continuously deformed to the map from a circle to a
single point; the maps belong to different homotopy classes. In the
non-Abelian case, this stability argument need no longer hold. With
more directions available in field space, there are more ways to
deform a vortex without leaving the vacuum manifold: if the latter
has the topology of a sphere rather than of a circle (i.e. if it is
simply connected), there is nothing preventing a loop on it to be
contracted to a single point, so classical stability of vortex
solutions is no longer guaranteed. Such solutions can still occur as
so-called embedded defects (solutions of a model based on a subgroup
of the symmetry group under consideration)
\cite{nambu1977}\cite{barriola1994} but to find out whether they are
stable, you must study their dynamics in detail. Since $\Phi$ going
to zero at some point is a common property of defects, whether
embedded or not, the unitary gauge is oblivious to their existence.

The problem is compounded by the non-trivial vacuum structure of YM
theory, even without Higgs. Consider the gauge transformation of Eq.
\eqref{nonAbelianWGaugeTransformation} applied to the trivial vacuum
solution $W_{a\mu} = 0$. Since a gauge transformation can not affect
observable quantities (e.g. energy density) the resulting ``pure
gauge'' configuration must be a vacuum solution too. But with a
non-Abelian symmetry group at our disposal, we can clearly use Eq.
\eqref{nonAbelianU} to create non-trivial maps between spacetime and
internal (i.e. group) space: the smallest non-Abelian Lie group,
SU(2), has three generators, so we have always at least one
generator for each spatial dimension. For instance, with
transformation parameters
\begin{eqnarray}\label{aMap}
\omega_a(x) = \frac{n \pi x_a}{\sqrt{x^2 + \kappa^2}}
\end{eqnarray}
(where $\kappa$ is an arbitrary number and $n$ is some integer) Eq.
\eqref{nonAbelianU} maps each point $\bm U$ in group space to $n$
points at spatial infinity ($x^2 \to \infty$). As with the Abelian
vortices, two maps with different winding number $n$ can not be
continuously deformed into each other, i.e. they belong to different
homotopy classes (unlike maps differing merely by $\kappa$).
Substituting them into Eq. \eqref{nonAbelianWGaugeTransformation}
therefore yields topologically distinct vacua, aptly called
$n$-vacua, separated by energy barriers in configuration space.
Since $n$ can be any integer, the vacuum of pure YM theory is
infinitely degenerate.

In the absence of other fields, transitions between $n$-vacua are
classically forbidden. Upon quantization, tunneling transitions
(instantons) become possible in principle, but remain far below
observable rates in practice
\cite{belavin1975}\cite{thooft1976}\cite{callan1976}\cite{jackiw1976}\cite{jackiw1976b}.
When you introduce $\Phi$ however, non-contractible loops
\cite{manton1983} and spheres \cite{klinkhamer1993} appear in
configuration space, implying the existence of saddle point (i.e.
unstable) static solutions perched on top of the barriers, along the
minimum-energy paths between neighboring $n$-vacua. Such solutions
are known as sphalerons (Greek for ``ready to fall''). Although they
were first discovered in the SU(2) \cite{dashen1974} and
U(1)$\times$SU(2) Higgs models
\cite{klinkhamer1984}\cite{yaffe1989}\cite{kleihaus2004} (where
Nambu's embedded vortices \cite{nambu1977} were also retroactively
recognized as sphalerons
\cite{klinkhamer1994}\cite{vachaspati1994}), they have since turned
up elsewhere too, from QCD \cite{mohapatra1992} to
Einstein-Yang-Mills theory (YM coupled to gravity)
\cite{bartnik1988}\cite{galtsov1991}, showing that the primary role
of $\Phi$ in enabling their emergence is to provide a finite energy
scale for the barriers. In the standard electroweak model, this
scale is $\sim 10 \thinspace TeV$.

The discovery of sphalerons posed a serious challenge to cosmology.
Transitions between electroweak $n$-vacua violate conservation of
baryon number (B), so in thermal equilibrium they erase any net
baryon number introduced as an initial condition or created at
higher energy scales (because the Boltzmann distribution gives equal
weight to particles of equal mass, hence to baryons and
anti-baryons). This is known as the baryon washout problem. In the
hot early universe, sphaleron transitions kept a B-violating channel
open all the way down to $T \sim 100 \thinspace GeV$, so
baryogenesis can not have happened much above the electroweak scale.
Fortunately, under non-equilibrium conditions sphalerons may also
provide an efficient mechanism for the production of a net B
\cite{kuzmin1985}\cite{trodden1998}\cite{dine2004}.

It should come as no great surprise that all these solutions feature
points where $\Phi = 0$, so they all live in the null space of the
unitary gauge. The problem has been known for a long time: gauge
groups are compact, but the unitary gauge insists on topological
triviality \cite{montonen1978}. A ``unitary gauge-like'' description
of all physical degrees of freedom is therefore necessarily singular
\cite{thooft1981}\cite{chernodub2008}.

\section{Quantum theory}

Brief mention of instantons and baryon number non-conservation
aside, everything we have done so far is strictly classical field
theory, i.e. our fields have been ordinary functions. But as Feynman
would say, ``I'm not happy with all the analyses that go with just
classical theory, because nature isn't classical, dammit''
\cite{feynman1982}. To make contact with the real world, we must
quantize. In quantum field theory (QFT) this means turning the
fields and their conjugate momenta into operators and imposing
canonical commutation relations between them. Observables are then
obtained as expectation values.

\subsection{Effective action}
Following Feynman, we introduce the action functional for a set of
fields $\Phi(x)$ coupled to sources $J(x)$ (all indices suppressed;
each field has its own source)
\begin{eqnarray}\label{theAction}
\langle {\cal L} + J \Phi \rangle = \int d^4x \left[{{\cal L}(x) +
J(x)\Phi(x)}\right]
\end{eqnarray}
and write the transition amplitude between the vacuum in the
infinitely far past and the vacuum in the infinitely far future as
the path integral
\begin{eqnarray}\label{WJ}
W[J] = \langle 0, +\infty | 0, -\infty \rangle = {\cal N} \int {\cal
D}\Phi \thinspace \textrm{e}^{i \langle {\cal L} + J\Phi \rangle}
\end{eqnarray}
where ${\cal N}$ is a normalization constant. The expectation value of
$\Phi$ can then be obtained by varying W[J] in J:
\begin{eqnarray}\label{dWdJ}
\frac{\delta W[J]}{\delta J(x)} = i \langle 0, +\infty | \Phi(x) | 0, -\infty \rangle_J
\end{eqnarray}
In practice it's more convenient to work with the functional Z[J]
(the generator of connected Green functions) satisfying
\begin{eqnarray}\label{ZJ}
W[J] = \textrm{e}^{i Z[J]}
\end{eqnarray}
The closest relative to a classical field is the semiclassical (or
``mean'') field
\begin{eqnarray}\label{semiclassicalPhi}
\Phi_{sc}(x) = \frac{\delta Z[J]}{\delta J(x)}
= \frac{\langle 0, +\infty | \Phi(x) | 0, -\infty \rangle_J}
       {\langle 0, +\infty | 0, -\infty \rangle}
\end{eqnarray}
In the absence of sources, it reduces to the vacuum expectation
value (VEV) of $\Phi(x)$; the condition for SSB in the quantum
theory is therefore
\begin{eqnarray}\label{quantumSSB}
\Phi_{sc}(x)_{J=0} \neq 0
\end{eqnarray}
If the functional dependence of $\Phi_{sc}(x)$ on $J(x)$ is
invertible, we can eliminate the explicit source dependence using
the functional Legendre transform
\begin{eqnarray}\label{effectiveAction}
\Gamma[\Phi_{sc}] = Z[J] - \langle J \Phi_{sc} \rangle
\end{eqnarray}
by which
\begin{eqnarray}\label{JofGamma}
J(x) = \frac{\delta \Gamma[\Phi_{sc}]}{\delta J \Phi_{sc}(x)}
\end{eqnarray}
Without interactions, this expression reduces to the classical
equations of motion for $\Phi$. With interactions, there are
corrections $\propto \hbar$ due to quantum fluctuations about the
classical trajectory.

$\Gamma[\Phi_{sc}]$ is known as the effective action. It can be
written as the multilocal expansion
\begin{eqnarray}\label{gammaMultiLocal}
\Gamma[\Phi_{sc}] &=& \sum^{\infty}_{n=1}\frac{1}{n!} \int d^4x_1
... d^4x_n \cdot \nonumber \\
&&\cdot\thinspace\thinspace \Phi_{sc}(x_1) ... \Phi_{sc}(x_n)
\Gamma^{(n)}(x_1, ..., x_n)
\end{eqnarray}
where $\Gamma^{(n)}(x_1, ..., x_n)$ are the proper vertices
(one-particle irreducible Green functions) of the theory. Taylor
expansion of $\Phi_{sc}(x_k)$ for $k > 1$ about $x_1$ yields the
quasilocal form
\begin{eqnarray}\label{gammaQuasiLocal}
\Gamma[\Phi_{sc}] = \langle
\frac{1}{2}F(\Phi_{sc})\partial^{\mu}\Phi_{sc}\partial_{\mu}\Phi_{sc}
- V(\Phi_{sc}) + ... \rangle
\end{eqnarray}
where the functions $F(\Phi_{sc})$ and $V(\Phi_{sc})$ summarize our
ignorance and ``...'' stands for higher order derivative terms.
$V(\Phi_{sc})$ is the effective potential. If we require the vacuum
to be translation invariant, the condition for SSB can be written
\begin{eqnarray}\label{effectivePotentialSSB}
\left. \frac{\partial V(\Phi_{sc})}{\partial \Phi_{sc}}
\right|_{\Phi_{sc} = \nu \neq 0} = 0
\end{eqnarray}
i.e. the effective potential must have an asymmetric minimum, just
like its classical counterpart.

Note that none of this involves perturbation theory. The shift of
Eq. \eqref{aGoldstoneVacuum} is sometimes motivated with an analogy
to elementary quantum mechanics, by anticipating the quantization of
small oscillations about the bottom of the potential and their
identification with particles, but as you can see, nothing in the
formalism requires it. You are free to quantize the original,
unshifted Lagrangian, and if you want to study non-perturbative
phenoma, like SSB, that is what you should do. The vacuum manifold
is then counterintuitively characterized by a non-vanishing
occupation number, i.e. the vacua are not empty, but rather filled
with the condensate of Eq. \eqref{quantumSSB}.

\subsection{Gauge fields}
When gauge fields are included, we are again confronted with the
redundacy which they bring about. The path integral in Eq.
\eqref{WJ} sums over all configurations, including those related by
gauge transformations, and so becomes even more divergent than
usual. The solution is to restrict the functional measure to only
one representative from each gauge-equivalent class (Faddeev-Popov
ansatz \cite{faddeev1967}). In practice, a factor $\det(M_f)
\delta(f(\chi))$ is introduced ($\chi$ now denotes both gauge and,
optionally, matter fields). The function $f(\chi)$ goes to 0 only
when the gauge condition is satisfied; $M_f = \delta f/\delta
\omega$ is the functional Jacobi matrix of $f(\chi)$ which encodes
its response to infinitesimal gauge transformations\footnote{See
\cite{kunstatter1992} and Section 9.1 in \cite{cheng1984} for an
introduction to the geometric interpretation of the Faddeev-Popov
ansatz in terms of a hyperplane which intersects each gauge orbit
once.}. The requirement that the gauge condition be
reversible\footnote{The Coulomb gauge famously fails to satisfy this
requirement at the non-perturbative level in the non-Abelian case
\cite{gribov1978}, and no gauge satisfies it when spacetime is a
four-sphere, i.e. the hypersphere of five-dimensional Euclidean
space \cite{singer1978}. While this may seem like a mathematical
curiosity, it's worth keeping in mind that which spacetime you're
working in can affect the validity of a gauge condition.} translates
to $\det(M_f) \neq 0$.

When $M_f$ depends on the fields, it's convenient to write
\begin{eqnarray}\label{detMf}
\det M_f = \textrm{e}^{\textrm{Tr}(\ln M_f)}
\end{eqnarray}
and treat the exponential as additional terms in the action. This is
done by introducing fictitious ``ghost'' fields. The full $\cal L$
going into Eq. \eqref{WJ} then consists of the original Lagrangian,
the ghost Lagrangian and $\delta(f(\chi))$ in exponential form.

A simpler alternative is to use a gauge condition with
field-independent $M_f$, so that $\det(M_f)$ can be trivially
factored out of the path integral. When the gauge condition is
sufficiently simple for the Lagrangian to be written directly in
terms of the independent variables, the $\delta(f(\chi))$ factor can
be explicitly enforced and so does not appear as a separate gauge
fixing term. A gauge with these properties is known as a physical
gauge, since it only uses physical degrees of freedom.

The axial gauges $W^a_0 = 0$ and $W^a_3 = 0$ are physical
gauges\footnote{The time-axial gauge $W^a_0 = 0$ is a borderline
case: it leaves a residual invariance under time-independent gauge
transformations which is removed by explicitly imposing Gauss' law.
Since it commutes with the Hamiltonian, this need only be done at
one point in time, e.g. on the initial conditions. See also Appendix
\ref{wildside}.} (see \cite{leibbrandt1987} for a review). The
naively defined unitary gauge is occasionally claimed to be one too,
but since it produces gauge bosons with mass terms on the same form
as those of the classical Maxwell-Proca Lagrangian for massive
vector fields, it succumbs to the same problem upon quantization: a
badly divergent perturbative expansion which can not be renormalized
and so is useless for (most) practical calculations. From a modern
perspective, it should come as no surprise that arbitrarily cutting
out a part of configuration space spoils renormalizability
(especially since the path integral runs over all field
configurations, whether classical solutions or not), but the
original work was done in the older canonical formalism.
Historically \cite{weinberg2004}, this is why the Weinberg-Salam
model of electroweak interactions was not taken seriously until 't
Hooft's proof of renormalizability in (initially a special case of)
the $R_\xi$ gauges \cite{thooft1971a}\cite{thooft1971b}.

The $R_\xi$ gauges provide the only way to make sense of the unitary
gauge at the quantum level that I know of. For the Abelian Higgs
model, the gauge condition is
\begin{eqnarray}\label{u1RxiGauge}
\partial_\mu A^\mu + \xi g\nu \phi_2 = 0
\end{eqnarray}
When it's imposed, the Goldstone boson acquires a mass squared $\xi
(g\nu)^2$. The $\xi$ parameter can take any value from 0 (Landau
gauge, classically equivalent to the Lorenz gauge, featuring
massless Goldstone bosons) through 1 (Feynman gauge, with Goldstone
and gauge boson of equal mass) to the formal limit $\xi \to \infty$.
The gauge boson propagator picks up an extra pole at the same mass
as the Goldstone boson, so propagating Goldstone bosons damp out
within a distance $\sim 1/(g\nu)$ \cite{fujikawa1972} (essentially
the same cancellation is seen in the axial gauge \cite{dams2004}).
This gives the figure of speech that gauge bosons ``eat'' Goldstone
bosons a much more direct interpretation than in the classical
theory: they really do! The classically obscure transformation of
Goldstone kinetic terms into gauge mass terms can now be understood
as a consequence of Goldstone bosons being confined within a Compton
wavelength of gauge bosons.

Gauge invariance implies that $\xi$ must drop out from any
calculation of physical observables, so leaving $\xi$ unassigned and
explicitly verifying that it does not enter the result provides a
powerful error check. To make contact with the unitary gauge, note
that the limit $\xi \to \infty$ is equivalent to making the
Goldstone bosons infinitely massive. Bringing an infinitely massive
particle into existence requires an infinite amount of energy, so
taking this limit effectively removes the Goldstone bosons from the
theory (at least as long as gravity is ignored). In this sense, the
limit $\xi \to \infty$ corresponds to the unitary gauge, so the two
are identified (a kind of argument known as ``physicist's
mathematics''). But note that this is really very different from
what we did in the classical theory.

The next time somebody tells you that ``Goldstone bosons are just
the longitudinal components of a massive gauge boson'', consider
asking if gauge bosons really have infinite mass. That might prove
entertaining.

\subsection{Nielsen identities}
Early work on SSB in QFT focused heavily on the effective potential,
$V(\Phi_{sc})$ in Eq. \eqref{gammaQuasiLocal}, and understandably
so: it is relatively easy to compute in a loop expansion (equivalent
to an expansion in powers of $\hbar$) and lends itself to an
intuitive interpretation as the quantum-corrected classical
potential. In the absence of gauge interactions, it is also easily
proved to be the expectation value of the energy density in the
lowest energy state satisfying the constraint that $\Phi_{sc}$ is
spacetime independent \cite{jona-lasinio1964}. Away from its minima,
$V(\Phi_{sc})$ generally has an imaginary component, signalling that
homogeneous field configurations are actually unstable, but this
inconsistency was often ignored.

Less easily ignored was Jackiw's observation that in gauge theories,
$V(\Phi_{sc})$ is gauge dependent \cite{jackiw1974}. Physical
quantities can not depend on the choice of gauge, so apparently
$V(\Phi_{sc})$ could not be an energy density in gauge theories. Not
in every gauge anyway. But maybe in a special one?

In \cite{dolan1974}, Dolan and Jackiw performed the change to polar
field coordinates (for the Abelian Higgs model) inside the path
integral (without the singularities retained in
\cite{chernodub2008}, so implicitly assuming the absence of
vortices), rewrote the resulting functional Jacobian as a ghost term
and argued that the resulting ``unitary Lagrangian'' is ``the unique
description of the physical dynamics of the system from which the
gauge degrees of freedom have been removed by a functional
integration''\footnote{This may well be how the meme that ``the true
dynamics is given by the unitary gauge'' was born.}. Happily, a
one-loop calculation of $V(\Phi_{sc})$ yielded a finite result which
could be reproduced in the $R_\xi$ gauges, provided that the unitary
limit $\xi \to \infty$ was taken before sending the momentum cutoff
in the loop integral $\to \infty$.

Of course, $V(\Phi_{sc})$ is by definition a static quantity, not a
dynamic one, and if there is any context where the unitary gauge
might be expected to yield a finite answer, despite the exclusion of
topologically non-trivial configurations, it should be one involving
spacetime independent configurations only. But the ``unitary
Lagrangian'' of \cite{dolan1974} is not the classical one: the new
ghost field term arising in the functional measure is now understood
to be equivalent to a quartically divergent, non-polynomial Higgs
self-coupling \cite{grosse-knetter1993}.

Even so, the suggestion that this ``unitary Lagrangian'' enjoys a
unique status was quickly put to rest by Nielsen \cite{nielsen1975}
and, independently, by Fukuda and Kugo \cite{fukuda1976}. Working in
the Fermi gauges
\begin{eqnarray}\label{fermiGauge}
f(A^\mu) = -\frac{1}{2\xi}\left(\partial_\mu A^\mu\right)^2
\end{eqnarray}
Nielsen showed that gauge invariance implies a simple differential
equation which relates the dependence of the effective potential on
$\xi$ and $\Phi_{sc}$, respectively:
\begin{eqnarray}\label{nielsenIdentity}
\left[\xi \frac{\partial}{\partial\xi} + C(\xi, \Phi_{sc})
\frac{\partial} {\partial\Phi_{sc}} \right] V(\xi, \Phi_{sc}) = 0
\end{eqnarray}
The function $C(\xi, \Phi_{sc})$ can be determined order by order in
the loop expansion. Essentially, this means that a variation in
$\xi$ is always compensated by one in $\Phi_{sc}$, keeping the value
of the effective potential invariant to each order of the expansion.
The generalization of this equation to the non-Abelian case and to
successively more general subclasses of the $R_\xi$ gauges was
carried out in
\cite{aitchison1984}\cite{johnston1984}\cite{nascimento1987}\cite{cima1999}.
The result is now collectively known as the Nielsen identities.

In \cite{fukuda1976}, a generalization of this approach was used to
show that the full effective action (effective potential plus
derivative terms) is gauge invariant at stationary points (so
solutions of the effective field equations are gauge invariant) to
each order of the loop expansion. The effective action can be
computed in any (workable) gauge, the energy density computed at its
stationary points is gauge invariant, the stationary points with the
smallest energy densities are the true vacua, and if the fields at
such a stationary point are not spacetime dependent, it follows that
their energy density is given by the effective potential. The last
condition is satisfied by a wide class of gauges (dubbed ``good
gauges'' by Fukuda and Kugo), including Coulomb, axial, Fermi and
$R_\xi$.

On the other hand, the effective action (and potential) is generally
\emph{not} gauge invariant away from its stationary points (``off
shell''). As first pointed out by Vilkovisky, this is due to the
dependence of the couplings between fields and external sources in
Eq. \eqref{theAction} on the chosen field parametrization
\cite{vilkovisky1984}. It is possible to write down Nielsen-like
identities which enforce invariance under field reparametrizations
and to construct effective actions which satisfy them, and which can
therefore be identified with energy density also off shell
\cite{burgess1987}. In particular, the so-called Vilkovisky-DeWitt
effective action for YM theory is just the ordinary effective action
evaluated in the covariant background field gauge
\cite{kunstatter1992}\cite{rebhan1987}: each field is written as the
sum $W^a_\mu + Q^a_\mu$ of a background part $W^a_\mu$ (neither
gauge fixed, coupled to a source or path integrated over) and of a
quantum part $Q^a_\mu$ (coupled to a source and path integrated over
as usual) whose covariant derivative with respect to $W^a_\mu$ is
required to vanish:
\begin{eqnarray}\label{backgroundGauge}
\partial_\mu Q_a^\mu + g C_{abc} W_{b\mu} Q_c^\mu = 0
\end{eqnarray}
Like all covariant gauge conditions, Eq. \eqref{backgroundGauge}
requires the introduction of ghosts. If on-shell gauge invariance is
all you need, an axial condition on $Q^a_\mu$ will do the job
without them.

\subsection{Finite temperature}
When reading old papers from the golden age of QFT, roughly mid-60s
to mid-70s, it is hard to miss the shifting view of gauge symmetries
and SSB: from neat mathematical tricks allowing the construction of
renormalizable theories to physical reality. The watershed event was
the realization that spontaneously broken symmetries are restored at
high temperature
\cite{kirzhnits1972}\cite{dolan1974b}\cite{weinberg1974}. As
Weinberg put it, ``if a gauge symmetry becomes unbroken for
sufficiently high temperature, then it is difficult to doubt its
reality'' \cite{weinberg1974}.

To see how this comes about, consider again the effective action.
Using the conventionally defined path integral of Eq. \eqref{WJ},
the proper vertices in Eq. \eqref{gammaMultiLocal} are the
vacuum-to-vacuum expectation values of time-ordered field operators:
\begin{eqnarray}\label{vacuumGreens}
\Gamma^{(n)}(x_1, ..., x_n) = \langle 0, +\infty | T[\Phi(x_1) ...
\Phi(x_n)] | 0, -\infty \rangle
\end{eqnarray}
In plain English, they describe sequences of scattering events
starting in empty space in the infinitely far past and ending in
empty space in the infinitely far future. If the system under study
is not empty space, this is not an adequate model. For instance, in
a heat bath in equilibrium at inverse temperature $\beta$, the
probability of a scattering event between energy eigenstates
$|\Phi_n\rangle$ with energy $E_n$ follows the Boltzmann
distribution
\begin{eqnarray}\label{boltzmann}
P_n = \frac{\textrm{e}^{-\beta E_n}}{\sum_m \textrm{e}^{-\beta E_m}}
\end{eqnarray}
Given a complete, orthonormal set of $|\Phi_n\rangle$, we should
then use the finite-temperature proper vertices
\begin{multline}\label{boltzmannGreensOrthogonal}
\Gamma_\beta^{(n)}(x_1, ..., x_n) = \\
\frac{\sum_l \textrm{e}^{-\beta E_l} \langle \Phi_l, +\infty |
T\left[\Phi(x_1) ... \Phi(x_n)\right] | \Phi_l, -\infty
\rangle}{\sum_m \textrm{e}^{-\beta E_m}}
\end{multline}
or, dropping the orthogonality requirement, the more general
\begin{eqnarray}\label{boltzmannGreens}
\Gamma_\beta^{(n)}(x_1, ..., x_n) = \frac{\textrm{Tr}
\textrm{e}^{-\beta H} T\left[\Phi(x_1) ...
\Phi(x_n)\right]}{\textrm{Tr} \textrm{e}^{-\beta H}}
\end{eqnarray}
where the trace runs over any complete set of states and $H$ is the
Hamiltonian. By comparison with the usual expression for the
transition amplitude from initial state $|\Phi_i\rangle$ to final
state $|\Phi_f\rangle$,
\begin{eqnarray}\label{transitionAmplitude}
\langle \Phi_f | \textrm{e}^{-t H} | \Phi_i \rangle = \int {\cal
D}\Phi \textrm{e}^{-\int_0^t d\tau \int d^3x {\cal L} }
\end{eqnarray}
we can therefore handle the (equilibrium) finite temperature case by
performing the formal substitution $t \to \beta$ and restricting the
functional integration to field configurations periodic in $\beta$
(antiperiodic for fermions)
\cite{matsubara1955}\cite{kubo1957}\cite{martin1959}. The generating
functional $W[J]$ of Eq. \eqref{WJ} then becomes the partition
function of statistical mechanics.

Using this technique to compute the effective potential
$V(\Phi_{sc})$ at finite temperature reveals that it picks up a
positive mass term $\propto 1/\beta^2$. At a sufficiently small
$\beta$ (the inverse critical temperature), this thermal mass term
becomes dominating and turns $\Phi_{sc} = 0$ into the global
potential minimum. The symmetry is then restored, i.e. $\Phi$ is
equally likely to be found anywhere on the zero temperature vacuum
manifold.

The argument that the ``unitary Lagrangian'' of \cite{dolan1974}
enjoys a unique status was dealt its first blow in this context (and
by its own originators). When the one-loop comparison with the
$R_\xi$ gauges was extended to finite temperature in
\cite{dolan1974b}, the critical temperatures did not match, and the
$R_\xi$ result was identified as the correct one.

The discrepancy was attributed to the non-renormalizability of the
unitary gauge, but this was counterintuitive: the compactification
of the time dimension which turns QFT into equilibrium thermal field
theory replaces loop integrals with sums over discrete frequency
modes, each described by its own three-dimensional theory, and lower
dimensional theories have better high energy behavior than higher
dimensional ones. Indeed, renormalization at zero temperature is
always sufficient to remove all infinities, so why did problems crop
up in the finite, thermal loop contributions and not in the zero
temperature part?

This became known as ``the unitary gauge puzzle''. The key to the
solution is the observation that while the value of the effective
potential is gauge invariant at stationary points (in ``good
gauges''), its curvature (used to obtain the effective Higgs mass in
\cite{dolan1974b}) is not: to compute the latter necessarily
involves going off shell \cite{ueda1980}, where the identification
of $V(\Phi_{sc})$ with energy density no longer holds. Even so, it
took several iterations
\cite{chaichian1991}\cite{arnold1992}\cite{kelly1994} (and a couple
of decades) to recognize that the correct critical temperature can
in fact be extracted order by order from any self-consistent
perturbative expansion, even (!) one based on the ``unitary
Lagrangian'' of \cite{dolan1974}.

Needless to say, the high temperature symmetric phase is in the null
space of the unitary gauge. There is also a problem common to all
gauges featuring unphysical degrees of freedom: the traces of Eq.
\eqref{boltzmannGreens} run over all of state space, but there is no
reason why ghosts etc. should be in thermal equilibrium with the
heat bath. The starting point for thermal gauge field theory is
therefore necessarily a physical gauge \cite{bernard1974}.

\subsection{Effective field theories}
The temperature dependence of $V(\Phi_{sc})$ is a reminder that
physics is not scale invariant; the world looks and acts differently
at different energy scales. Fortunately, you don't need detailed
knowledge of physics at the scales of grand unification or quantum
gravity to understand physics at the electroweak scale, you don't
need detailed knowledge of physics at the electroweak scale to
understand nuclear physics, and you don't need detailed knowledge of
nuclear physics to understand chemistry.

In QFT, this independence of lower energy phenomena from higher
energy ones is formally known as the Appelquist-Carazzone decoupling
theorem \cite{appelquist1975}. What it says is that massive fields
effectively decouple at low energy: given a renormalizable
Lagrangian containing both massless and massive fields, you can
describe its low energy behaviour with a renormalizable Lagrangian
written in terms of the massless fields only. The massive fields
only contribute to the low energy Lagrangian through the
renormalization of its couplings and fields.

More generally, you can eliminate heavy fields from a Lagrangian
which also contains light fields by encoding their effects in
(generally non-renormalizable) interaction terms involving the light
fields only. The resulting effective field theory (EFT) is valid at
energies below the masses of the eliminated fields
\cite{georgi1994}. The whole edifice of standard model extensions --
technicolor, supersymmetry, grand unified theories (GUTs),
supergravity and string theory with its infinite tower of massive
excitations -- implicitly depends on the suppression by powers of
energy over mass of the effective interaction terms induced by the
high energy extensions.

In principle, constructing an effective field theory from a more
fundamental one is straightforward. Take the effective action of Eq.
\eqref{gammaMultiLocal}, Fourier transform it and do all momentum
integrals involving the heavy fields. Split the remaining integrals
in two parts, one for momenta going up to your cutoff (the mass of
the lightest eliminated field or less), one for momenta above the
cutoff, and do the latter too (alternatively, gauge-fix and
integrate out the heavy fields only, leaving any local invariance
specific to the light fields unbroken until the need actually arises
to gauge-fix them too \cite{weinberg1980}). Transform back, and you
are left with a non-local action on the form of Eq.
\eqref{gammaQuasiLocal}, written in terms of the light fields only
and valid down to distance scales $\sim$ 1/(momentum cutoff).

In practice, this ``top-down'' program may not be possible to carry
out, even approximately, either because you don't know the
fundamental Lagrangian or because there is no workable approximation
scheme. In such cases, you may still be able to create an EFT by
systematically writing down all interaction terms (up to some cutoff
dimension) involving light fields only and respecting all known
symmetries. The schoolbook example of this approach is chiral
perturbation theory of low energy QCD, a perturbative expansion in
masses and momenta of quark bound states, small on the hadron mass
scale $\sim 1\thinspace GeV$
\cite{weinberg1979}\cite{gasser1984}\cite{gasser1985}. In this case,
the fundamental Lagrangian is known, but ordinary perturbation
theory breaks down at low energy due to the growth of the effective
coupling (confinement).

Fortunately, this problem does not occur in Higgsed YM theories:
small couplings stay small also at low energy, and the
Appelquist-Carazzone theorem ensures the existence of an EFT written
in terms of the massless fields only (if any). Since corrections to
the tree level terms are $\propto$ powers of couplings over energy,
integrating out the massive gauge bosons from a weakly coupled
theory induces negligible effects at energies $\ll$ the symmetry
breaking scale. In this low energy/long distance limit, the EFT
reduces to the classical Lagrangian \emph{with all massive gauge
bosons set to zero}\footnote{I emphasize this because I have
occasionally noticed some confusion on this point, especially among
cosmologists, who sometimes seem to take classical Lagrangians a
little too literally.}.

More generally, the classical theory should provide a good
approximation to the finite temperature dynamics of modes with large
occupation number, i.e. for fields with mass $\ll$ temperature
\cite{grigoriev1988}\cite{felder2001}\cite{rajantie2002}.

\subsection{Non-linear sigma models as EFTs}
Consider the low energy limit of the original U(1) $\sim$ O(2)
Goldstone model. A look at Eqs.
\eqref{nlGoldstone}-\eqref{nlMexPotential} is enough to tell that
the effective Lagrangian must be
\begin{eqnarray}\label{lowEnergyGoldstone}
{\cal L}(\theta) = \frac{1}{2}\nu^2 \partial_\mu\theta \partial^\mu\theta
\end{eqnarray}
i.e. that of a free, massless scalar field (the Goldstone boson)
taking values on a circle in internal space (since $\theta$ is an
angle, so $\theta = 0$ and $\theta = 2\pi$ are identified; the
circle is of course the bottom of the Mexican hat potential). There
can't be any $\rho$ particles on shell when the energy density is
$\ll$ the rest mass of a $\rho$ per Compton volume, i.e. $\sim
\lambda^2\nu^4$, and quantum corrections at energy $E$ can't be
worse than $\sim \lambda^2 E/\nu$.

The factor $\nu^2$ in front of the derivative term is the
``stiffness'' of $\theta$. Make it larger and it costs more energy
to lift $\theta$ from the ground state (an arbitrarily chosen,
constant $\theta$ value). Note that massless is not synonymous with
``cheap''; when we speak of ``low energy EFT'', we really mean ``low
energy \emph{density} EFT'' (just like we really mean ``Lagrangian
density'' when we say ``Lagrangian'' in field theory). This is why
the derivative term of a massless field belongs in a low energy EFT
even if it is associated with a very large energy scale: unlike a
mass term, a derivative term has a continuous spectrum (unless
Lorentz symmetry is broken, e.g. by a spatially periodic potential
\textit{a la} Kronig-Penney) so arbitrarily small values are
allowed.

There is more interesting physics in Eq. \eqref{lowEnergyGoldstone}
than meets the eye, all due to the identification of $\theta = 0$
with $\theta = 2\pi$. As in the Abelian Higgs model, topology
guarantees the existence of vortex solutions, but without a matching
gauge field, the gradient energy of a single vortex diverges
logarithmically with spatial radius. This divergence can be cured by
pairing up a vortex with an anti-vortex winding in the opposite
direction, so that their far fields cancel. The vortices in such a
pair attract with a logarithmic potential; add more pairs and you
can create a periodic vortex lattice. A two-dimensional ``gas'' of
vortices and anti-vortices also features a high temperature
transition above which thermal disorder wins over pairwise
attraction and individual vortices can roam freely
(Kosterlitz-Thouless transition).

Things get even more interesting when you consider the
generalization of the Goldstone model to higher O(N) groups. The low
energy limit is then given by the Lagrangian
\begin{eqnarray}\label{generalSigmaModel}
{\cal L}(\vec{n}) = \frac{1}{2}\nu^2 \partial_\mu\vec{n} \cdot
\partial^\mu\vec{n}
\end{eqnarray}
where $\vec{n}$ is an N-dimensional vector constrained to take
values on the (N - 1)-sphere, i.e. satisfying $|\vec{n}| = 1$.
Mathematicians call this a wave map, physicists a non-linear sigma
model (NLSM).

For N = 3, $\vec{n}$ is a 3-vector taking values on an ordinary unit
sphere (a 2-sphere). In terms of standard spherical coordinates
$(\vartheta, \varphi)$,
\begin{eqnarray}
n_1 &=& \sin(\vartheta) \cos(\varphi) \\
n_2 &=& \sin(\vartheta) \sin(\varphi) \\
n_3 &=& \cos(\vartheta)
\end{eqnarray}
and the Lagrangian reads
\begin{eqnarray}\label{sphericalO2L}
{\cal L}(\vartheta, \varphi) = \frac{1}{2}\nu^2 \left[
\partial_\mu\vartheta \partial^\mu\vartheta +
\sin^2(\vartheta) \partial_\mu\varphi \partial^\mu\varphi \right]
\end{eqnarray}
Add a quartic derivative term and you get either the Skyrme model
\cite{skyrme1961}, a precursor to modern chiral perturbation theory,
or the Skyrme-Faddeev model \cite{faddev1975}, a candidate low
energy EFT for pure SU(2) YM theory. The extra terms are introduced
to defeat Derrick's theorem, a simple scaling argument which rules
out the existence of finite-energy static solutions to scalar
theories with quadratic gradient terms in more than two spatial
dimensions \cite{derrick1952}. On its own, the NLSM can only have
time-dependent three-dimensional solutions with finite energy.

Note the non-linear term (the ``NL'' in NLSM), which suggests richer
dynamics than in O(2), but also note that the O(2) case is recovered
when either $\varphi$ or $\vartheta$ is constant. This is easily
seen to generalize: higher O(N) models embed lower ones.

The construction of O(3) NLSM solutions is greatly simplified by the
stereographic projection to the complex plane (extended with a point
at infinity to represent the north pole)
\begin{eqnarray}\label{stereographicProjection}
u(\vartheta, \varphi) = \tan(\vartheta/2)\textrm{e}^{i\varphi}
\end{eqnarray}
after which
\begin{eqnarray}\label{cp1Lagrangian}
{\cal L}(u) = 2\nu^2 \frac{\partial_\mu u^\dag
\partial^\mu u}{(1 + u^\dag u)^2}
\end{eqnarray}
In this form, the model is known as CP$^1$ (Complex Projective, one
dimension). Writing $u = p + i q$ with real $p$ and $q$, the
equations of motion are
\begin{eqnarray}
\partial_\mu \partial^\mu p + 2 p \frac{\partial^\mu q \partial_\mu q - \partial^\mu p \partial_\mu p}{1 + p^2 + q^2} &=& 0 \\
\partial_\mu \partial^\mu q + 2 q \frac{\partial^\mu p \partial_\mu p - \partial^\mu q \partial_\mu q}{1 + p^2 + q^2} &=& 0
\end{eqnarray}
If $u$ is analytic in the $(x, y)$ plane, the Cauchy-Riemann
equations guarantee that $p$ and $q$ satisfy both the Laplace
($\nabla^2 p = 0$) and the eikonal ($(\nabla p)^2 = 0$) equation and
therefore also the equations of motion in $(x, y)$. A separable
factor which satisfies the wave equation along $z$, i.e. $u(t \pm
z)$, also works, so you can easily build either static solutions
constant in $z$ (e.g. $u = (x + i y)^n$, a vortex winding $n$ times
around the $(x, y)$ plane) or wave packets of arbitrary
three-dimensional shape moving up and down the $z$ axis.

But the real news relative to the O(2) case is the existence of
nontrivial maps from the 2-sphere in field space to the 2-sphere at
spatial infinity, as in
\begin{eqnarray}\label{o3hedgehog}
u &=& \left(\frac{x + i y}{r + z}\right)^n \\
r &=& \sqrt{x^2 + y^2 + z^2}
\end{eqnarray}
(see \cite{wereszczynski2005} for more). When $n = 1$, this is known
as a hedgehog. In line with Derrick's theorem, its energy is
proportional to volume, but like the global vortex, it has a gauged
counterpart with finite energy: the monopole first described by 't
Hooft and Polyakov \cite{thooft1974}\cite{polyakov1974}.

The next step up the NLSM ladder is O(4), i.e. a 4-vector taking
values on a unit 3-sphere. Its new feature is the existence of
knot-like maps from the internal 3-sphere to the spatial 2-sphere
(``textures'' to cosmologists). Unlike vortices and hedgehogs, those
do not have a singular core corresponding to $\Phi_{sc} = 0$ in the
full Goldstone model. Derrick's theorem guarantees that they too
must be unstable: they shrink until their gradient energy density
becomes too large for the low energy EFT to handle. In the full
model, the collapse ends with $\Phi_{sc}$ sliding over the top of
the potential, allowing the knot to unwind. This never happens given
suitably ``small'' initial conditions, however \cite{liebling2000}.
Topologically trivial configurations (sometimes referred to as
``non-topological textures'') can therefore keep evolving
indefinitely.

Hadron physicists know all about the O(4) NLSM. When the vector
components are light mesons (bound pairs of up and down quarks), it
corresponds to the leading order terms of chiral perturbation
theory. The ``sigma'' in ``sigma model'' comes from the nuclear
isospin singlet which acquires a VEV (the ``Higgs particle''); the
three independent vector components (the Goldstone bosons) are the
pions. Vortices, hedgehogs and even polyhedral maps have all been
described in this context, and it has been suggested that bubbles of
a Disoriented Chiral Condensate (DCC) rotated away from its value in
our QCD vacuum may be possible to create in collider experiments
\cite{cornwall1979}\cite{nelson1987}\cite{anselm1989}\cite{bjorken1991}\cite{blaizot1992}\cite{huang1996}\cite{ioannidou2004}.

In three spatial dimensions, the map from internal 3-sphere to
spatial 2-sphere is the last one with separate homotopy classes, so
higher O(N) NLSMs do not add qualitatively new kinds of solutions to
those found at N = 4. This also implies that the O(4) NLSM can
provide a reasonable approximation to the dynamics of the general N
$>$ 4 case \cite{turok1991}.

\section{Cosmology}

If you subscribe to standard hot big bang cosmology, the relevance
of high temperature symmetry restoration is obvious. Immediately
after the bang, the universe was filled with a high temperature
plasma in near thermal equilibrium. As it cooled, $V(\Phi_{sc})$
gradually changed shape, from harmonic to quartic, and $\Phi_{sc}$
eventually rolled or tunnelled from $\Phi_{sc} = 0$ to a randomly
picked, asymmetric minimum\footnote{This does not imply that the
finite temperature effective potential provides a good description
of the dynamics of the transition from symmetric to broken symmetry
phase. $V(\Phi_{sc})$ is computed for spacetime independent
$\Phi_{sc}$ and (at finite temperature) using the equilibrium
partition function. Spacetime independent, equilibrium
configurations only occur at extrema of the potential, so
$V(\Phi_{sc})$ can only be used to determine from which critical
temperature a phase transition is allowed, not to study its
dynamics. See \cite{boyanovsky1996} and \cite{felder2001}.}. Given a
particle horizon, i.e. a finite maximum radius within which light
(or anything else) could have traveled, this phase transition could
not be coordinated over all space, so causally disconnected regions
picked vacua independently of each other.

\subsection{Domains and dark energy}
This mechanism was recognized already by Weinberg in his seminal
paper on high temperature symmetry restoration \cite{weinberg1974},
which introduced the now familiar analogy with domain formation in
ferromagnetism and went on to ask: ``Does the universe consist of
domains, in which symmetries are broken in equivalent but different
directions? If so, what happens when a particle or an observer
travels from one domain to another?''

The second part of Weinberg's question was quickly (but, as it turns
out, only partially) answered by A. Everett, who considered the fate
of light crossing between domains in the standard electroweak model
\cite{everett1974} and found it to depend on the width of the
boundary: ``If the transition between domains is smooth, with a
transition region whose thickness is large compared with a
wavelength of the incident radiation, there is no reflection and one
observes a transmitted wave identical to the incident wave.'' A thin
boundary, on the other hand, would result in total reflection. After
the discovery in 1998 of a luminosity deficit from high redshift
Type Ia supernovae (SNe Ia)
\cite{riess1998}\cite{perlmutter1998}\cite{perlmutter1999}, his
description of the intermediate case, a domain boundary of width
comparable to wavelength, seems almost prophetic: ``A distant object
viewed through such a domain boundary would appear less bright than
it should. If one knew independently the distance and brightness of
the object, say by knowing its cosmological red-shift and the
brightness of objects similar to it, then the existence of such
semitransparent domain walls would be detectable.'' This was
published in 1974!

But there is a problem with Everett's derivation which would not
have been lost on Feynman: it's a purely classical treatment of the
electroweak boson sector (linearized to boot). The decay to fermions
of virtual weak bosons traversing a wide boundary is not considered.
For a macroscopic boundary width and an incident photon packing
enough energy to create at least a neutrino pair, the negligible
lifetime $\sim 10^{-25} \thinspace s$ of weak bosons (and the
absence of reflection at tree level) clearly makes this the
dominating effect. The problem then reduces to projecting the photon
(i.e. massless) state in the source domain onto the photon state in
the destination domain, and the residual luminosity for domains
related by SU(2) parameters $\vec{\omega} = [\omega_1, \omega_2,
\omega_3]$ is easily found to be \cite{anderberg2005}
\begin{eqnarray}\label{defl}
\ell({\vec{\omega}}) = \sin^2(\theta_W) \frac{\omega_\bot^2
\cos(|\vec{\omega}|) + \omega_3^2}{|\vec{\omega}|^2} +
\cos^2(\theta_W)
\end{eqnarray}
where $\omega_\bot^2 = \omega_1^2 + \omega_2^2$ and $\theta_W$ is
the Weinberg angle, $\sin^2{\theta_W} \simeq 0.2216$. One or two
electroweak domain boundaries between us and high redshift SNe Ia
would explain their luminosity deficit without new physics, and may
also provide the missing energy density required by the apparent
flatness of the universe, so obviously everybody but me hates the
idea.

Besides desperation for any sign of new physics, why did nobody
follow up Everett's work in 1998? A possible explanation is that it
had simply been forgotten\footnote{To answer the inevitable
question: no, I was not aware of it when I wrote
\cite{anderberg2005}, nor does Penrose seem to have been when he
made the remarks \cite{penrose2004} which got me thinking about this
problem.}. Another is the popularity of inflation: when the
supernova news broke, the first thought to cross the mind of most
physicists probably was ``Oh, so it's still going on!'' (it
certainly was my first thought).

Yet another reason might have been Everett's use of the term
``domain wall'', which had since become synonymous with
``topologically stable two-dimensional defect passing through
$\Phi_{sc} = 0$''. The standard electroweak model notoriously fails
to satisfy Kibble's group-theoretic criteria for the existence of
domain walls \cite{kibble1976}, and so must any sensible alternative
to it, since a domain wall network would quickly and
catastrophically collapse to a massive, single wall dominating its
horizon volume \cite{harvey1982}\cite{vachaspati1984}\footnote{A
related problem is posed by topologically stable monopoles, which
occur in any gauge theory featuring SSB of a simple group to a
semisimple group containing a U(1) factor, like the standard
electroweak group. This was one of the original selling points of
inflation: if the standard model descends from a simple GUT by SSB,
the energy density of a single GUT monopole per horizon volume at
the symmetry breaking transition would overclose the universe by
many orders of magnitude. Inflation solves the problem by diluting
monopoles and other defects left over from GUT symmetry breaking to
acceptable levels.}.

In hindsight, it may seem strange that \emph{transients} (i.e. not
necessarily stable configurations) interpolating between different
values of $\Phi_{sc}$ \emph{on} the vacuum manifold (rather than
passing through $\Phi_{sc} = 0$) were not considered in this
context. That they existed in the early universe, and may have
played an important cosmological role as seeds of large-scale
magnetic fields \cite{diaz-gil2008}, was already well understood
\cite{vachaspati1991}\cite{enqvist1993}\cite{kibble1995}\cite{copeland1996}\cite{hindmarsh1998}\cite{tornkvist1998}.
But by now you should have no problem guessing the (erroneous)
argument against their survival to late times: transforming such
configurations to the unitary gauge turns them into collections of
weak gauge bosons, which are unstable and decay in a microphysical
time\footnote{Yes, I was once summarily dismissed with this argument
by a supposedly top journal.}.

One way to see that this can not be right is to recognize the
implicit assumption that only the Higgs vacuum manifold is
degenerate. The transformation to the unitary gauge would then leave
the gauge bosons with only one vacuum to ``fall'' to, so no
coordination across space would be needed to complete the relaxation
to \emph{the} vacuum. That this assumption is incorrect in the
standard electroweak model has been elegantly demonstrated by Lepora
and Kibble, who showed that the vacuum manifold of the electroweak
gauge bosons is actually a squashed hypersphere \cite{lepora1999}.
The unitary gauge conceals this degeneracy in a ``spaghetti code''
mess of interaction terms which I won't even try to reproduce here
(see e.g. Ch. 14 in \cite{mandl1984}).

More generally, even disregarding that Goldstone bosons are not
``just longitudinal components of massive gauge bosons'' in a
consistently quantized theory, it is obvious from the vantage point
of a physical gauge that the argument must be wrong, since it
assigns a privileged role to an arbitrarily chosen point on the
vacuum manifold. It is isomorphic to somebody in New York claiming
that people in London are not standing vertically and so must all be
falling over: after all, they are not aligned with the Empire State
Building\footnote{An esteemed editor at another top journal recently
informed me that the unitary gauge can always be applied using
partially overlapping patches. This is isomorphic to claiming that
the existence of a complete world atlas made of partially
overlapping, flat charts proves that Earth is flat. Yes,
flat-earthers are funny.}!

Since Earth's (idealized) surface and the vacuum manifold are both
Riemannian and symmetric, it is of course always possible to connect
any two points on either one by a curve featuring a continuous
sequence of tangent spaces, each having equivalent \emph{but
different} definitions of ``vertical'' (or of ``gauge boson X'').
Your ability to stand vertically at your current location is
sufficient to know that you could create a chain of people
connecting any two cities on Earth, all standing vertically
according to their own, local definition. Likewise, the existence of
a stable gauge boson in a single vacuum is sufficient to know that
you could create interpolating configurations between any two points
on the vacuum manifold using only stable gauge bosons. Massless
gauge bosons are necessarily stable (below some fermion pair
production threshold \cite{schwinger1951}\cite{shabad2006}), so if
your theory has them, you are all set. Nambu gave us a convenient
sufficient criterion for their existence: the broken symmetry group
must be semisimple (as in the standard electroweak model) or have
rank $\geq 2$ \cite{nambu1977}. In other words, the microphysical
decay argument fails in any realistic theory.

Microphysical stability does not imply classical stability of
macroscopic configurations, of course. What it does imply is that
the realignment of $\Phi_{sc}$ after the initial, random choice of
vacua in causally disconnected regions must be resolved according to
the effective field equations. In a cosmological model with finite
particle horizons, this is enough to guarantee the existence of
electroweak (and maybe other) domains. Their boundaries, whether
classically stable or not, might be possible to inflate beyond the
current horizon with a late period of exponential expansion, but
that would require new physics at the electroweak scale (or lower)
and risk diluting away any net baryon number created at the
electroweak phase transition.

\subsection{Global and local textures}
In the early 90s, textures, topological \cite{turok1989} and not
\cite{phillips1995}, were considered serious candidate sources of
the primordial density perturbations which seeded the large scale
structure of the universe. More recently, it has been suggested that
the large anomalous cold spot found by WMAP in the cosmic microwave
background (CMB) is due to a texture which originated at a symmetry
breaking scale $8.7\times 10^{15}\thinspace GeV$ (intriguingly close
to the hypothetical GUT symmetry scale $\sim 10^{16}\thinspace GeV$
\cite{langacker1993}) and eventually collapsed at redshift $z \sim
6$, a billion years after the big bang \cite{cruz2007}.

The textures in question have all been global, of the O(N $>$ 3)
Goldstone model $\to$ NLSM variety. This runs counter to the
commonly held view that continuous symmetries should be local (i.e.
gauged), and it runs into serious difficulties with quantum gravity
(string theory in particular does not seem to allow any global
continuous symmetries; see e.g. p. 255 in \cite{kiritsis2007}). The
reason for nevertheless sticking with global textures is the
following argument, due to Turok \cite{turok1989}:

A scalar field $\Phi$ constrained to the vacuum manifold can be
written
\begin{eqnarray}\label{phiOnManifold}
\Phi(x) = {\bm U}(x) \Phi_0
\end{eqnarray}
where ${\bm U}(x)$ is a symmetry transformation and $\Phi_0$ is an
arbitrarily chosen reference point on the manifold. If $\Phi$ is
gauged, with covariant derivative ${\bm D}_\mu$ given by Eq.
\eqref{nonAbelianCovariantDerivative}, the gauge field can ``fall''
to
\begin{eqnarray}\label{pureGaugeW}
{\bm W}_\mu = \frac{i}{g} \left(\partial_\mu {\bm U}\right){\bm
U}^{-1}
\end{eqnarray}
at every point in space. This makes ${\bm D}_\mu \Phi = 0$, so the
gradient energy of $\Phi$ is zero. Since ${\bm W}_\mu$ is a pure
gauge configuration, its energy also vanishes and the configuration
stops evolving. Naively, this relaxation process should complete on
the time scale of the gauge interaction, i.e. in a microphysical
time, so any gauged texture will be gone long before it can have
observable consequences.

It is not hard to see where this argument came from. If we start
from the trivial vacuum $\Phi(x) = \Phi_0$, ${\bm W}_\mu(x) = 0$ and
apply the gauge transformation ${\bm U}(x)$ according to Eqs.
\eqref{nonAbelianPhiTransformation} and
\eqref{nonAbelianWGaugeTransformation}, we obtain Eqs.
\eqref{phiOnManifold} and \eqref{pureGaugeW}. A gauge transformation
can not affect physical observables, so the energy density must
remain zero.

In other words, ${\bm U}(x)$ is just the inverse of the
transformation used to impose the unitary gauge, so this is just the
usual microphysical decay argument in light disguise. How does it go
wrong? Let me count the ways.

\textbf{One}: As often pointed out by Vachaspati, there is no reason
why $\Phi$ should have vanishing gradient energy at finite
temperature. By dimensionality and equipartition of energy alone,
given an inverse temperature $\beta$ we should expect ${\bm D}_\mu
\Phi \sim 1/\beta^2$.

\textbf{Two}: At first sight it would seem that this argument must
work for any field configuration. But you already know that domain
walls, Nielsen-Olesen vortices and 't Hooft-Polyakov monopoles are
stable. Clearly, the argument fails for them. Why? An easily seen
reason is that they have cores where $\Phi(x) = 0$, which can not be
written according to Eq. \eqref{phiOnManifold} (i.e. they are in the
null space of the unitary gauge). Try setting ${\bm U} = 0$ and the
${\bm U}^{-1}$ in Eq. \eqref{pureGaugeW} blows up. Stated another
way, ${\bm U}$ fails to be unitary at the core of such a defect:
${\bm U}^\dag \neq {\bm U}^{-1}$. This clearly does not apply to
textures, which have no singular core, but topological textures
nevertheless owe their existence to the existence of maps in
different homotopy classes. Transformations from one homotopy class
to another are by definition discontinuous, so even if the ${\bm
U}^{-1}$ in Eq. \eqref{pureGaugeW} won't blow up, the $\partial_\mu
{\bm U}$ will. Gauging the theory lets you move the discontinuity
from $\Phi$ to ${\bm W}_\mu$, but does not eliminate it.
%Maybe add footnote here pointing out that in other words, the argument
%implicitly assumes that $\Phi$ and ${\bm W}_\mu$ both start out in the
%same homotopy class, or that there is enough thermal energy to overcome
%the 10 TeV barrier separating them by a sphaleron transition.

\textbf{Three}: Even topologically trivial $\Phi$ configurations
carry conserved quantities (energy, momentum, gauge currents, all in
derivative terms) which must go elsewhere, i.e. to fermions, upon
relaxation. Fermions can only be produced effectively while the
energy and charges within the Compton volume of a fermion pair (e.g.
electron + anti-neutrino for electroweak interactions) are $\geq$
the total mass and charges of such a pair (on shell). Once $\Phi$
gradients fall below this threshold, dissipation to fermions becomes
exponentially suppressed. The evolution of $\Phi$ and ${\bm W}_\mu$
then becomes a Hamiltonian flow, but need not stop.

Combine the first and third point and you can estimate the initial
size beyond which charged $\Phi$ configurations should have been
safe from dissipation: the horizon scale when $\beta$ crossed above
the Compton length of the lightest charged fermion. For electroweak
interactions, that's $\sim 10^3 \thinspace km$ \cite{anderberg2005}.
If such configurations then simply tracked overall metric expansion,
their corresponding minimum size today would be $\sim$ one
lightyear. Apply the same redshift to their energy density and you
inevitably land right at the current ``dark energy'' scale,
$10^{-3}\thinspace eV$.

\textbf{Four}: As Nambu pointed out long ago, ${\bm D}_\mu \Phi = 0$
does \emph{not} guarantee vanishing energy when there are linear
combinations of commuting generators which annihilate $\Phi$, i.e.
when massless gauge fields remain after SSB \cite{nambu1977}.

For a heuristic understanding of this point, consider the standard
electroweak model, which conveniently has the same scalar sector as
the O(4) Goldstone model: setting the photon $A_\mu(x) = 0$ along
with the massive gauge bosons and applying some ${\bm U}(x)$ will
certainly yield a gauge-equivalent vacuum according to Eqs.
\eqref{phiOnManifold} and \eqref{pureGaugeW}. But thanks to the
absence of a mass term, any finite, constant $A_\mu$ is an equally
valid vacuum. Starting from such a configuration, i.e. from ${\bm
W}_\mu \propto 1 + \tau_3$ (see \cite{anderberg2005}) Eq.
\eqref{nonAbelianWGaugeTransformation} picks up an additional term
not contained in Eq. \eqref{pureGaugeW}. The common term does not
depend on the gauge fields, so evidently the same ${\bm U}(x)$
yields different vacua for different choices of initial constant
$A_\mu$. This is a simple example of the degeneracy of the gauge
sector's vacuum \cite{lepora1999}. The implication is that just as
for the scalars, an independent choice of vacuum at each point in
space will not generally result in a global energy minimum. That
requires coordination, so the usual causality bound (the horizon)
applies.

In a slightly more mathematical language, for any ``good gauge'' (in
the sense of Fukuda and Kugo \cite{fukuda1976}) the lowest energy
configurations are spacetime independent, so the energy density in a
sufficiently small neighborhood of such a minimum (in field space)
can be approximated by an ordinary Taylor expansion. Since there are
no linear terms, the first non-trivial term is the Hessian matrix of
second order derivatives in the fields, i.e. the mass matrix. For
the minimum to be non-degenerate, the discriminant (the determinant
of the Hessian) must be positive definite. If there is at least one
massless field, this condition is not satisfied: the minimum is
degenerate and SSB ensues.

This can be seen explicitly by working out the low energy EFT of the
standard electroweak model in an axial gauge \cite{anderberg2007}.
You can start from scratch or you can get a head start by using the
gauged NLSM (GNLSM), valid for energies $\ll$ the Higgs mass (now
known to be $> 100\thinspace GeV$), which was written down long ago
to parametrize the unknown symmetry breaking sector
\cite{appelquist1980}\cite{longhitano1980}\cite{longhitano1981} (see
Ch. 2 in \cite{kilian2003} for a review). Choosing the second route,
the Goldstone field matrix in polar field coordinates $\theta_a$ is
\begin{eqnarray}\label{su2Sigma}
{\bm \Sigma} = \textrm{e}^{i \theta_a \tau_a / 2} = \cos(\theta/2) +
i\thinspace \frac{\theta_a\tau_a}{\theta}\sin(\theta/2)
\end{eqnarray}
with $\theta = \sqrt{\left(\theta_1\right)^2 +
\left(\theta_2\right)^2 + \left(\theta_3\right)^2}$, and
\begin{eqnarray}\label{phiThetaEquivalence}
\left[
\begin{array}{lr}
\phi^{0\dag} & \phi^+\\
-\phi^{+\dag} & \phi^0
\end{array}
\right] = \frac{\nu}{\sqrt{2}} {\bm \Sigma}
\end{eqnarray}
The covariant derivative acting on ${\bm \Sigma}$ is
\begin{eqnarray}\label{DSigma}
{\bm D}_\mu = \partial_\mu + i \frac{g_W}{2} W^a_\mu \tau^a - i
\frac{g_B}{2} B_\mu \tau^3
\end{eqnarray}
and the full Lagrangian is
\begin{eqnarray}\label{polarLagrangian}
{\cal L}_{GNLSM} &=& -\frac{1}{4}B_{\mu\nu}B^{\mu\nu} - \frac{1}{4}W^a_{\mu\nu}W^{a\mu\nu} \nonumber \\
&&+ \frac{\nu^2}{4} {\rm Tr}\left[\left({\bm D}_\mu {\bm
\Sigma}\right)^\dag \left({\bm D}^\mu {\bm \Sigma}\right)\right]
\end{eqnarray}
You can easily convince yourself that it is equivalent to the
standard electroweak boson Lagrangian with the radial Higgs degree
of freedom clamped to its VEV (just write out all terms explicitly
and compare). It has been used as is to compute the one-loop thermal
effective action for an electroweak plasma at temperatures below the
mass of a (heavy) Higgs and above that of the weak gauge bosons
\cite{manuel1998}, but we are interested in the real low energy
limit, where the massive gauge bosons can not be excited either and
must be set to zero along with radial Higgs excitations (so there is
no assumption about the Higgs mass here, other than that it is much
larger than the interaction energies under consideration). We
therefore need to isolate the linear combination of gauge fields
with zero mass, i.e. the photon, as defined in an arbitrary vacuum
$\vec{\theta} = [\theta_1, \theta_2, \theta_3]$.

In the basis $\left[B_\mu, W^1_\mu, W^2_\mu, W^3_\mu\right]$, the
${\cal L}_{GNLSM}$ terms quadratic in $B_\mu$ and $W^a_\mu$ give
rise to the mass matrix
\begin{eqnarray}\label{gaugeMassMatrix}
\frac{\nu^2}{2}
\left[
\begin{array}{cccc}
g_B^2 & g_B g_W \Theta_1 & g_B g_W \Theta_2 & -g_B g_W \Theta_3 \\
g_B g_W \Theta_1 & g_W^2 & 0 & 0 \\
g_B g_W \Theta_2 & 0 & g_W^2 & 0 \\
-g_B g_W \Theta_3 & 0 & 0 & g_W^2
\end{array}
\right]
\end{eqnarray}
where we have introduced the convenient auxiliary quantities
\begin{eqnarray}
\Theta_1 &=& \left[\theta_1\theta_3(\cos(\theta)-1)+\theta\theta_2\sin(\theta)\right]/\theta^2 \label{Theta1} \\
\Theta_2 &=& \left[{\theta_2\theta_3}(\cos(\theta)-1)-{\theta\theta_1}\sin(\theta)\right]/\theta^2 \label{Theta2} \\
\Theta_3 &=& \left[(\theta_1^2+\theta_2^2)\cos(\theta) + \theta_3^2\right]/\theta^2 \label{Theta3}
\end{eqnarray}
satisfying $\left(\Theta_1\right)^2 + \left(\Theta_2\right)^2 +
\left(\Theta_3\right)^2 = 1$. The eigenvalues of Eq.
\eqref{gaugeMassMatrix} are the tree level masses squared of photon,
$W^{\pm}$ and $Z^0$. The two degenerate eigenstates can be
orthogonalized to obtain\footnote{Note that the scalar product of
Eq. \eqref{eigenA} (properly normalized) for $\vec{\theta} = 0$ and
$\vec{\theta} = \vec{\omega}$ is Eq. \eqref{defl}.}
\begin{eqnarray}
A_\mu           &\propto& [g_W/g_B,  -\Theta_1,          -\Theta_2,    \thinspace\thinspace\Theta_3 ] \label{eigenA} \\
\acute{W}^1_\mu &\propto& [0, \quad\quad\thinspace\thinspace-\Theta_2, \quad\Theta_1, \quad 0       ] \label{eigenW1} \\
\acute{W}^2_\mu &\propto& [0,         \Theta_1 \Theta_3, \Theta_2 \Theta_3, \Theta_1^2 + \Theta_2^2 ] \label{eigenW2} \\
Z_\mu           &\propto& [-g_B/g_W, -\Theta_1, -\Theta_2, \Theta_3]
\label{eigenZ}
\end{eqnarray}
To eliminate the massive bosons, invert Eqs.
\eqref{eigenA}-\eqref{eigenZ} and set $\acute{W}^1_\mu =
\acute{W}^2_\mu = Z_\mu = 0$ (note that given orthogonal
eigenstates, inversion amounts to normalizing them, assembling them
in a column matrix and transposing). The result is
\begin{eqnarray}
B_\mu &=& A_\mu \cos(\theta_W) \label{bFromA} \\
W^1_\mu &=& -A_\mu \Theta_1 \sin(\theta_W) \label{w1FromA} \\
W^2_\mu &=& -A_\mu \Theta_2 \sin(\theta_W) \label{w2FromA} \\
W^3_\mu &=&  A_\mu \Theta_3 \sin(\theta_W) \label{w3FromA}
\end{eqnarray}
To obtain the low energy EFT of the electroweak boson sector to
leading order, substitute Eqs. \eqref{bFromA}-\eqref{w3FromA} into
${\cal L}_{GNLSM}$ and find
\begin{eqnarray}\label{masterLagrangian}
{\cal L}_{EFT} &=& \frac{\nu^2}{8}\left[ \partial_\mu \theta \partial^\mu \theta
                                  + \frac{4 \sin^2(\theta/2)}{\theta^2}
                                    \left( \frac{\vec{\theta}}{\theta} \times \partial_\mu \vec{\theta}\right)^2 \right] \nonumber \\
&&- \frac{1}{4} \left( \partial_\mu A_\nu - \partial_\nu A_\mu \right)^2 \nonumber \\
&&- \frac{\sin^2(\theta_W)}{4} \left( A_\mu \partial_\nu \Theta_a - A_\nu \partial_\mu \Theta_a \right)^2
\end{eqnarray}
The first row in Eq. \eqref{masterLagrangian} is just the plain O(4)
NLSM in polar field coordinates, the second row is the Maxwell
Lagrangian, the third row couples them $\propto \sin^2(\theta_W)$,
acting as an effective photon mass term when $\vec{\theta}$ is not
constant. Everett's classical result is now easily recovered by
noting that a photon can penetrate a region of varying
$\vec{\theta}$ only if its total energy exceeds its effective mass,
which is $\propto$ the $\vec{\theta}$ gradient.

For any constant $\vec{\theta}$, ${\cal L}_{EFT}$ reduces to plain
electromagnetism, as it must\footnote{A truly stunning objection
raised against ${\cal L}_{EFT}$ is that it is less symmetric than
the full electroweak Lagrangian, so ``important terms are missing''.
Yes, Virginia, the low energy EFT has less symmetry than the full
theory! It's called SSB, you may have heard of it.}. Note that $\nu
\simeq 246.3\thinspace GeV$ makes the electroweak vacuum \emph{very}
stiff; you would need $\sim 10^{41}$ joule to ``melt'' a cubic
centimeter of it and create the electroweak equivalent of a DCC. In
cold war terms, that's $\sim 10^{25}$ megatons of TNT, enough to
completely vaporize Earth, journals included, a couple billion
times. Being immersed in our vacuum, such a configuration would also
immediately snap back to it (unless dynamically stable, which might
be theoretically possible \cite{graham2007}). It is only on
astronomical scales that random $\vec{\theta}$ gradients can be
shallow enough to be long-lived.

Specializing to the time-axial gauge\footnote{Even if you are
uncomfortable with the time-axial gauge at the quantum level, this
is not a problem after the EFT has been obtained using e.g. the
covariant background field gauge. In the present case this is a
non-issue, since we are staying at tree level and disregarding
quantum corrections as negligible at low energy.} ${\bm W}_0 = 0$,
the electric and magnetic fields are
\begin{eqnarray}
 \vec{E} &=& -\partial_0\vec{A} \\
 \vec{B} &=& \nabla \times \vec{A}
\end{eqnarray}
Varying ${\cal L}_{EFT}$ in $A_0$ yields the Gauss constraint
\begin{eqnarray}\label{gauss}
\nabla \cdot \vec{E} = \partial_0 \vec{\theta}^T {\bm H} \vec{A}
\cdot \nabla \vec{\theta}
\end{eqnarray}
(a plane in $\vec{A}$ space) where the matrix ${\bm H}$ has
components
\begin{eqnarray}\label{H}
H_{ab} = \sin^2(\theta_W) \frac{\partial\Theta_c}{\partial\theta_a}
\frac{\partial\Theta_c}{\partial\theta_b}
\end{eqnarray}
This completely fixes the gauge, leaving $\vec{A}$ with only two
independent degrees of freedom. The energy density becomes a
manifestly non-negative sum of quadratic forms,
\begin{eqnarray}\label{rho}
\rho &=& \frac{1}{2}\left(\vec{E}^2 + \vec{B}^2\right) \nonumber \\
&&+ \frac{1}{2} \partial_0\vec{\theta}^T \left(\nu^2 {\bm G} + \vec{A}^2 {\bm H} \right) \partial_0\vec{\theta} \nonumber \\
&&+ \frac{\nu^2}{2} \partial_m \vec{\theta}^T {\bm G}
\partial_m \vec{\theta} \nonumber \\
&&+ \frac{1}{2} \left(\varepsilon_{jkl} A_k \partial_l
\vec{\theta}\right)^T {\bm H} \left(\varepsilon_{jmn} A_m \partial_n
\vec{\theta}\right)
\end{eqnarray}
where ${\bm G}$, with components
\begin{eqnarray}\label{G}
G_{ab} = \left(\frac{\delta_{ad}\delta_{be}}{2} + \frac{1 -
\cos{\theta}}{\theta^2} \varepsilon_{cda}\varepsilon_{ceb} \right)
\frac{\theta_d\theta_e}{\theta^2}
\end{eqnarray}
is the 3-sphere metric, with eigenvalues $1/4$ and (doubly
degenerate) $0 \leq (1-\cos(\theta))/(2\theta^2) \leq 1/4$. ${\bm
H}$ also has no negative eigenvalues (but zeros along all axes of
$\vec{\theta}$ space)\footnote{Note the emergence of two natural
metrics in field space, one ($G_{ab}$) associated with the scalar
sector and one ($H_{ab}$) with the gauge sector, in line with
\cite{lepora1999}.}, so by the spectral theorem, $\vec{A} \neq 0$
can only increase $\rho$ for a given $\vec{\theta}$.

Since the low energy EFT of electroweak interactions (valid whether
the minimal model is the correct UV completion or not) contains the
O(4) NLSM, any solution obtained in the latter, textures included,
is also a valid electroweak solution on macroscopic scales. It
should be obvious that this result generalizes to larger symmetry
groups. In particular, any viable GUT must embed the standard model
and so must contain at least nine massless gauge fields (eight
gluons and a photon, but there may be more; ``hidden'' sectors which
only couple to the standard model fields via gravity are commonplace
e.g. in string phenomenology).

Usual unitary gauge shenanigans aside, the one intelligent concern
which has been raised in this context, echoing the discussion in
\cite{anderberg2007}, is that the coupling between photon and
$\vec{\theta}$ (the third row in Eq. \eqref{masterLagrangian}) may
significantly affect the dynamics of electroweak textures,
invalidating numerical simulation results obtained in the plain O(4)
NLSM. While only simulations based on the full EFT will tell for
sure, it should be noted that GUTs tend to have complicated Higgs
sectors (even the original, minimal SU(5) model needs two Higgs
multiplets, one in the adjoint to obtain the standard model group,
one in the fundamental representation to break electroweak symmetry)
and that the NLSM of a GUT could have viable subgroups which do not
couple directly to the photon. To the extent that the O(4) NLSM is a
passable representative of the general N $\geq$ 4 case, it should
also be a reasonable approximation of the dynamics of textures
arising in such subgroups.

Finally, I can't resist pointing out that the kind of energy
released by the collapse of an electroweak texture could conceivably
match that of even the largest gamma ray bursts (GRBs), $\sim
10^{47}$ joule (equivalent to one cubic meter of electroweak scale
plasma). That may be something worth thinking about when confronted
with ``shots in the dark'' not easily accomodated by the standard
massive stellar collapse model, like GRB 070125, and one more reason
to undertake dynamic simulations of ${\cal L}_{EFT}$.

\section{Conclusion}

The unitary ``gauge'' is strictly speaking not a gauge, but an
unphysical constraint which excludes an important part of
configuration space and conceals the full symmetry of the theory. It
is oblivious to non-perturbative solutions like sphalerons and
topological defects, even at the classical level, and is undefined
in the high temperature symmetric phase. Upon quantization, it
yields a non-renormalizable and therefore mostly useless
perturbative expansion. Identifying it with the singular limit $\xi
\to \infty$ of the $R_\xi$ gauges, equivalent to making Goldstone
bosons infinitely massive, saves renormalizability, but only at the
cost of introducing unphysical degrees of freedom which can not be
assumed to be in thermal equilibrium with physical ones.

All these difficulties can be avoided by adopting a physical gauge,
as is routinely done in non-perturbative numerical simulations.
Carrying out the effective field theory program in such a gauge
makes the transition from symmetric high temperature phase to broken
low temperature phase explicit and suggests that several high
profile problems in cosmology may have a common denominator: the
overly simplified picture of gauge symmetry breaking painted by the
unitary gauge.

\appendix
\section{Off the beaten path}\label{wildside}
This review presents the mainstream, semiclassical picture of the
Higgs mechanism: at low energy, the massless modes of the classical
Lagrangian are believed to provide a good first approximation to the
non-perturbative background, on top of which minor quantum
corrections can be computed. There are other views.

As I write this, no Higgs particle has ever been detected. Even if
something like the standard electroweak Higgs boson is eventually
found, you could still take the stance, sometimes advocated by
Veltman, that the perturbative expansion \emph{defines} the theory,
while the electroweak Lagrangian and path integral are nothing more
than convenient ``bookkeeping devices'' used to aid in the
construction and evaluation of Feynman diagrams \cite{veltman2000}.
If this is the case, there is no such thing as a non-perturbative
electroweak sector.

The strongest reason not to believe this is the success of lattice
QCD. The lattice is essentially a discretization of the full path
integral; QCD is a YM theory, like the standard electroweak model;
and on the lattice, QCD reproduces the light hadron mass spectrum to
better than 1\% \cite{durr2008}. This is a result from deep
non-perturbative territory which no summation of Feynman diagrams
could ever reproduce. If the path integral works so well beyond
perturbation theory for QCD, why should it not work beyond
perturbation theory for electroweak interactions?

But what the lattice giveth, the lattice can taketh away.

There is an argument, based on lattice regularization, by which no
quantity with vanishing mean value on its gauge orbit can have a
non-zero VEV. While the original demonstration, known as ``Elitzur's
theorem'' \cite{elitzur1975}, falls well short of what
mathematicians would call a theorem (it actually shows that a
compact Abelian gauge field minimally coupled to an O(2) sigma model
can not have a finite VEV when the lattice is taken to the continuum
limit), the argument is commonly believed to generalize
\cite{splittorff2003}.

The main thrust is (almost) obvious by inspection: explicitly, the
path integral expression for the expectation value $\langle Q
\rangle$ of some $\Phi$-dependent quantity $Q(\Phi)$ is
\begin{eqnarray}\label{expectationValue}
\langle Q \rangle = {\cal N} \int {\cal D}\Phi \thinspace Q(\Phi)
\thinspace \textrm{e}^{i \langle {\cal L} + J\Phi \rangle}
\end{eqnarray}
The exponentiated action is not affected by symmetry
transformations, so for each orbit around field space (e.g. the
bottom of the Mexican hat potential) it can be factored out of the
integral, reducing the integrand to
\begin{eqnarray}
Q(\Phi) \thinspace \textrm{e}^{i \langle {J\Phi \rangle}}
\end{eqnarray}
For vanishing external source ($J = 0$), if $Q(\Phi)$ averages to
$0$ over an orbit, the integral over that orbit will vanish; if this
holds for all orbits, $\langle Q \rangle$ must also vanish. The hard
part is proving that $\langle Q \rangle \to 0$ for finite $J \to 0$,
which is not immediately obvious since the path integral runs over
an infinite-dimensional space\footnote{From a cosmological
perspective, nitpicking about finite ``external'' sources may strike
you as nonsensical until you realize that as far as you are
concerned, everything now entering your particle horizon is an
external source, quantum and thermal fluctuations about a vanishing
expectation value included.}.

Putting the path integral on the lattice turns it into a finite
number of nested, ordinary integrals, one for each lattice site and
field. Taking the continuum limit then reveals that the argument
fails for global symmetries (i.e. $\langle Q \rangle$ can remain
finite for arbitrarily small, finite $J$), but apparently does hold
for local symmetries\footnote{A major loophole is that the continuum
limit is unlikely to exist for theories which are not asymptotically
free. Theories with a U(1) product group, like the standard
electroweak model, are not; that's one reason to view them as low
energy EFTs derived from something more fundamental. If the
underlying theory is a simple GUT, the continuum limit may exist. If
the underlying theory is something else, maybe involving a
fundamental length scale, all bets are off.}. In particular, this
implies that the vacuum expectation value of a Higgs field must
vanish: $\Phi_{sc} = \langle\Phi\rangle = 0$.

What should we make of this? Elitzur's own assessment was clear: for
SSB to occur, the gauge symmetry must be explicitly broken by
imposing a gauge condition which leaves some global symmetry
unbroken (either the gauge symmetry restricted to spacetime
independent transformations or some other symmetry). It is the
global symmetry that is spontaneously broken. This is in line with
the standard (well informed) interpretation of the Higgs mechanism,
and with the classical view that the theory is undefined until the
gauge is completely fixed.

Remember the ``good gauges'' of Fukuda and Kugo \cite{fukuda1976}:
only after a gauge condition is imposed can you know whether a given
field configuration is a vacuum. The usual textbook presentation
does things the other way around; first a ``vacuum'' is picked, then
the unitary gauge is imposed. The validity of this procedure can
only be verified \textit{a posteriori}, and as we have seen, it does
not pass the check at the non-perturbative level. Indeed, the
unitary gauge and the $R_\xi$ gauges (for generic $\xi$) break the
gauge symmetry both locally and globally, unlike most common gauges.
Neither is therefore adequate for studying SSB (as opposed to
studying perturbations on a given, constant Higgs background, a.k.a.
particle physics).

So far so good. But what if you impose a gauge condition and
$\langle\Phi\rangle$ still vanishes?

In \cite{creutz1978} it was argued that the residual invariance of
the time-axial gauge under time-independent gauge transformations
implies (for the Abelian Higgs model, in the canonical operator
formalism) $\langle\Phi\rangle = 0$. This result was obtained for
``physical'' states of infinite norm, which may give you some pause
(are field expectation values well-defined?), but the same
conclusion was reached again by a different route in
\cite{frohlich1981}, for arbitrary symmetry groups, using the
lattice-regularized path integral (in Euclidean space, subject to
the usual provisos about the validity of analytic continuation to
Minkowski spacetime and the existence of a continuum
limit)\footnote{Keep in mind that $\langle\Phi\rangle = 0$ does not
imply $\Phi = 0$. Quantum fluctuations alone guarantee that
$\langle\Phi^\dag \Phi\rangle$ can not vanish, even in the symmetric
phase.}.

If $\Phi$ were a physical observable, this would mean that repeated
independent measurements must yield random values with average 0 and
variance $\langle\Phi^\dag \Phi\rangle$, and we would have to
conclude that $\Phi$ is in a superposition state smeared out
symmetrically over each gauge orbit (an idea first floated in
\cite{fischler1975}). But since $\Phi$ is an unobservable, gauge
dependent quantity, all we can say (if we believe this result) is
that $\langle\Phi\rangle$ (and $\langle\Phi^\dag \Phi\rangle$) is
not a good order parameter in the time-axial gauge, i.e. that it is
not useful for the determination of the presence and nature of SSB
\emph{in that particular gauge} (in \cite{fischler1975}, it was
argued that $\langle\Phi\rangle$ is a good order parameter in the
Landau gauge only).

While the vanishing of $\langle\Phi\rangle$ in the time-axial gauge
may thus be nothing more than a quirk of that (incomplete) gauge
condition, in \cite{frohlich1981} it was conjectured to be a
\emph{physical} disorder effect caused by a ``gas'' of instantons
(and maybe other topologically non-trivial configurations) of
microphysical size, with average separation on the order of the
inverse Higgs mass. Elitzur notwithstanding, the gauge symmetry
would then also remain unbroken under global transformations.

In the electroweak context, there is an obvious problem with this
picture: contrary to observation, neither photons (as ordinarily
defined) nor left-handed fermions would propagate freely on everyday
distance scales, which are $\gg$ the inverse Higgs mass
(right-handed fermions would, since they are weak isospin singlets).
The only way out of this dilemma is another conjecture, first
publicized in \cite{banks1979} but attributed to Susskind: all
fields which are not singlets under weak isospin, i.e. the SU(2)
subgroup of the standard electroweak group $U(1)\times SU(2)$, are
confined to singlet bound states, like quarks and gluons under color
SU(3) of QCD. What we see as freely propagating particles are either
such singlet bound states or fundamental singlets. In particular,
the left-handed fermions observable at low energy all consist of a
fundamental left-handed fermion bound to a fundamental Higgs boson.
The observable ``Higgs particle'' is actually a bound pair of
fundamental Higgs + anti-Higgs.

The confinement conjecture was, once again, motivated by lattice
results. When simple models with Higgs fields in the fundamental
representation were put on the lattice and their parameters were
varied (in the ``frozen Higgs'', i.e. GNLSM approximation), no
indication was found of a phase boundary (discontinuities in
physical quantities) separating the confinement regime (small Higgs
VEV, large gauge coupling) and the Higgs regime (large Higgs VEV,
small gauge coupling) \cite{fradkin1979}. No qualitative difference
therefore seems to exist between the two regimes, in the same sense
that no qualitative difference exists between water vapor and liquid
water\footnote{Subsequent work showed that this property, dubbed
``complementarity'', can be lost when the radial Higgs degree of
freedom is allowed to fluctuate \cite{damgaard1985} and when
fermions are included \cite{lee1987}\cite{hsu1993}. These days,
invoking complementarity therefore requires additional assumptions
about unknown non-perturbative dynamics, in particular the absence
of spontaneous chiral symmetry breaking.}. But there are significant
quantitative differences, first detailed in \cite{abbott1981b}.

For the confinement conjecture to work, the non-Abelian interaction
must be strongly coupling. Electromagnetism is not strongly
coupling, so there can be no significant weak gauge boson
contribution to the photon, i.e. no significant mixing between U(1)
and SU(2) gauge bosons. The electric charges of observable particles
are then just their U(1) hypercharges, and the photon does not
interact significantly with the SU(2) instanton gas. Since the three
massive gauge bosons are an almost pure SU(2) triplet, they must
have almost identical masses, significantly larger than the standard
ones due to the strong SU(2) coupling ($\sim 125 \thinspace GeV$).
At low energy, the particle spectrum is the same as in the standard,
weakly coupled electroweak model, but the composite nature of quarks
and leptons starts showing up around $80 \thinspace GeV$, the mass
of the standard $W^\pm$ bosons. At higher energies, there is a
complicated spectrum of hadron-like bound states.

This scenario was all but killed in 1983 by the experimental
detection of the $W^\pm$ bosons at $80 \thinspace GeV$ and of the
$Z^0$ boson at $91 \thinspace GeV$, as predicted by the standard,
weakly coupled electroweak model. Subsequently, precision
electroweak measurements have shown no sign of composite structure
up to energies well in excess of $100 \thinspace GeV$, also ruling
out attempts to save electroweak confinement through additional
assumptions about its non-perturbative dynamics \cite{sather1996}.

Backtracking, the demise of the electroweak confinement conjecture
reduces the instanton gas suggestion of \cite{frohlich1981} to an
interesting thought experiment. It may still be relevant to other
theories, but not to the only example of the Higgs mechanism
actually known (?) to be realized in nature.

What's left? Out on the fringes of the arXiv, you can still
occasionally see (presumably well-meaning) arguments that SSB is
impossible in QFT on fundamental quantum mechanical grounds, because
a symmetric Hamiltonian can not produce an asymmetric state starting
from a symmetric one. By the same logic, it is clearly impossible to
find out whether Schr\"{o}dinger's cat is dead or alive, or to
measure the $J_x$ angular momentum component of an electron known to
have $J_z = 1/2$. You may also be told that a wave function prepared
on one side of a double well potential will evolve to a symmetric
shape, and that by analogy, classically stable topological defects
can unwind without passing through energetically forbidden
configurations, courtesy of quantum tunneling. If that happens,
inquire about the rate of such events in realistic field theories,
then compare the answer with the age of the universe.

Yet another argument which you may come across is that the universe
can not be likened to open thermodynamic systems, where SSB
demonstrably occurs, because unlike such systems it is isolated and
thus not subject to random external disturbances. I don't know about
the universe of those who write such things, but the universe which
I can see around me is a ball which has been growing at the speed of
light for the past 14 billion years (give or take a few), and for
all that time, its boundary in every direction has been sampling a
heat bath hot enough to melt the electroweak vacuum, and who knows
what else.

That works for me.

\section{Book tips}
Georgi once started off a nice review of effective field theories
\cite{georgi1994} with the following piece of advice about old
literature on almost any subject: ``Ignore it! With rare exceptions,
old papers are difficult to read because the issues have changed
over the years.'' I respectfully disagree. There is often much more
to the original papers than what you'll find in streamlined textbook
presentations. Read them, and you may discover hidden or forgotten
gems.

But textbooks obviously have their place, and there are some which
you may find particularly useful if field theory is not your day
job. Since I am an old guy, I know mostly old books. Don't worry
though. You could have slept through the past quarter century
without missing much if any new particle theory worth knowing about.

If you are going to get only one book, get Cheng \& Li
\cite{cheng1984}. It starts with basic QFT, thoroughly works through
the whole standard model and does not stop before it has also told
you about technicolor and SU(5). The introductions to SSB and to
path integral quantization of gauge theories are excellent. Journal
editors should read them religiously at least once every full moon.

The most accessible QFT textbook I know of remains Ryder
\cite{ryder1985}. Among other things, it features the best
introduction to non-perturbative aspects.

Another excellent QFT textbook which also covers the standard
electroweak model and SU(5) is Bailin \& Love \cite{bailin1986}.
Unusually for a textbook, it introduces finite temperature field
theory and high temperature symmetry restoration. It is also a good
place to start learning about anomalies, which you need to
understand if you want to understand the role of sphalerons in
baryogenesis.

More advanced discussions of path integral quantization, SSB and
field theory at finite temperature can be found in Rivers
\cite{rivers1987}.

For an introduction to supersymmetry and string theory, or just for
something more recent, consider Dine \cite{dine2007}, but understand
that it's a \emph{very} compressed presentation of essentially all
fundamental physics, the kind of book best enjoyed when you already
know the subject. If you don't, see it as a starting point for your
literature searches.

Hat off to Dine for being secure enough in his physicisthood to
openly admit that he finds symmetry breaking ``a puzzling notion in
gauge theories'' (p. 17).
%\newline\newline

\begin{acknowledgments}
I thank Michael Creutz and Roman Jackiw for suggested readings.
Feedback is welcome from friends, foes and neutrals alike who might
wish to inform me of typos, hair raising blunders (especially on my
part) or impending assassination attempts. Requests to remain
anonymous will of course be honored.
\end{acknowledgments}

\bibliography{basename of .bib file}

\end{document}